\documentclass[graybox, nosecnum]{svmult}

\usepackage{mathptmx}       
\usepackage{helvet}         
\usepackage{courier}        
\usepackage{type1cm}        
\usepackage{makeidx}         
\usepackage{graphicx}        
\usepackage{multicol}        
\usepackage[bottom]{footmisc}
\usepackage{hyperref}        
\usepackage{soul}            

\makeindex             

\begin{document}
\title*{The Fission Barrier of Heaviest Nuclei From a Macroscopic-Microscopic
Perspective}
\author{Michał Kowal\thanks{corresponding author} and Janusz Skalski }
\institute{Michał Kowal \at National Centre for Nuclear Research, Pasteura 7, 02-093 Warsaw, Poland, \email{michal.kowal@ncbj.gov.pl}
\and Janusz Sklaski \at National Centre for Nuclear Research, Pasteura 7, 02-093 Warsaw, Poland, \email{janusz.skalski@ncbj.gov.pl}}
%
%
\maketitle
\abstract
 {The concept of fission barrier - a parameter which enters in quantitative
  estimates of various observables related to nuclear fission -
   is presented from the point of view of theory based on the picture of
   nuclear deformation and energy dependent on it.
   We describe the macroscopic-microscopic method of calculating energy
   landscapes which is simpler than the selfconsistent mean field approach,
   and, due to its two-component nature, seems to be easier to adjust to
    experimental data. We present some models and methods used to find the
   fission saddles. For the purpose of illustration, we present results
  on fission barriers in actinides and superheavy nuclei, obtained
  within one macroscopic-microscopic model. We discuss comparisons
  with results of other models, including some of the mean-field type.  }

\section{Introduction}

One of the most interesting subjects in the contemporary nuclear physics is
 the existence and stability
of the heaviest atomic nuclei sometimes called superheavy.
Since nuclear fission is one of their dominant decay channels, determining
the main characteristics of this process is indispensable to estimate
their half-lives.

 Nearly immediately after its discovery, nuclear fission, both induced
 by neutrons \cite{Hahn1939} and spontaneous
 \cite{Petrzhak1940}, was interpreted in analogy to the process
 of cell division in biology \cite{Meitner1939}.
 Atomic nucleus was represented as a drop of liquid \cite{Bohr1939,Bohr21939},
 after \cite{Gamow1930}.
 The process itself was imagined as an increase in deformation under the
 disruptive electrostatic repulsion that, in spite of the counteracting
 nuclear attraction, eventually leads to scission.
 The notion of the fission barrier came about by imagining potential energy
 surface resulting from those forces and understood as a function of
 shape, or deformation $q$, of a drop of "nuclear matter". The energy surface
 should have a minimum at the deformation corresponding to the ground state
 (g.s.) of a nucleus and show a barrier along any path connecting it with a
  configuration of two nascent fragments, beyond which the potential should
  decrease with their distance, see Fig. \ref{schem}. Along any path
 connecting the ground state with one of possible scission points there would
 be a point of maximal energy, $V_{max}$.
  The minimum over all such paths of the differences $V_{max}-E_{g.s.}$,
 with $E_{g.s.}$ the ground state energy, is the fission barrier height $B$,
\begin{equation}
 \label{barr1}
  B ={\rm min}\ _{\rm paths}\{V_{max} - E_{g.s.}\} .
\end{equation}
 It corresponds to some $V(q_{\rm saddle})$ where
 $q_{\rm saddle}$ designates deformation of the saddle point.

 Theoretical models giving energy of a nucleus as a function of its
 shape belong to the selfconsistent mean-field variety, either the
 Hartree-Fock (HF) or Hartree-Fock-Bogolyubov (HFB - when accounting for
 pairing correlations)
 formalism using some effective nucleon-nucleon (usually density-dependent)
 interaction or a density functional (if there is no underlying Hamiltonian).
 There exist also a relativistic alternative of the mean field theory in the form of
 Relativistic Mean Fields or Relativistic Hartree-Bogolyubov formalism.
 The fundamental quantities in such theories are single-particle neutron
 and proton densities which are constructed from eigenstates (orbitals)
 of the single-particle mean-field Hamiltonian whose eigenvalues are the
  single-particle energies. The HF(B) procedure, usually by means of energy
 minimization, gives one solution -
 the ground state or some other local energy minimum. A way to obtain
 the energy landscape consists in applying constraints when minimizing energy
 - energy map arises from many repeated calculations with varying constraints.
 The one-body densities provide deformation or shape parameters through
 their spatial moments: quadrupole, octupole, hexadecapole
 etc., defined in the center-of-mass frame of the nucleon density. At present,
 such theories are the most advanced of practically possible ones for the
 heaviest nuclei.
 They are covered by the N. Schunck contribution to this volume.

The purpose of this chapter is to familiarize the reader with a hybrid but quite effective method that allows estimating/determining the fission barrier $B$ for heavy and superheavy nuclei.
 It is called the Macroscopic - Microscopic method (MM), as the calculation is
 divided into two parts:
 the macroscopic part involves most of the binding energy which has a smooth
 variation with $Z$ and $N$; the microscopic part involves a sum of individual
 nucleon contributions, showing oscillations with $Z$ and $N$ coming from
 the shell structure.

First, we present a general view on the fission process and our general
 understanding of it according to the excitation energy available to the
 decaying system, which points to the importance of the fission barrier.
Then we present the shell correction method in connection to the mean-field
 theory.
We discuss some models describing the macroscopic energy and
 phenomenological single-particle potentials.
 The numerical diagonalization of the s.p. Hamiltonian including this
 potential delivers the s.p. spectra,
 to which the shell correction method can be applied.
 Then we discuss the selection of collective variables which build deformation
 space used in calculations and specific difficulties of a search for saddles
 in multidimensional spaces.
Finally, the results of the MM method concerning fission barriers in
 actinides and superheavy nuclei are illustrated on the example of
 one specific model. We provide also some comparisons to experimental
 data and results of other MM and selfconsistent models.

\section{General view of the fission process - the importance of the fission barrier}

\begin{figure}[h]
\vspace{-0mm}
\hspace{4mm}
\centerline{\includegraphics[scale=0.35]{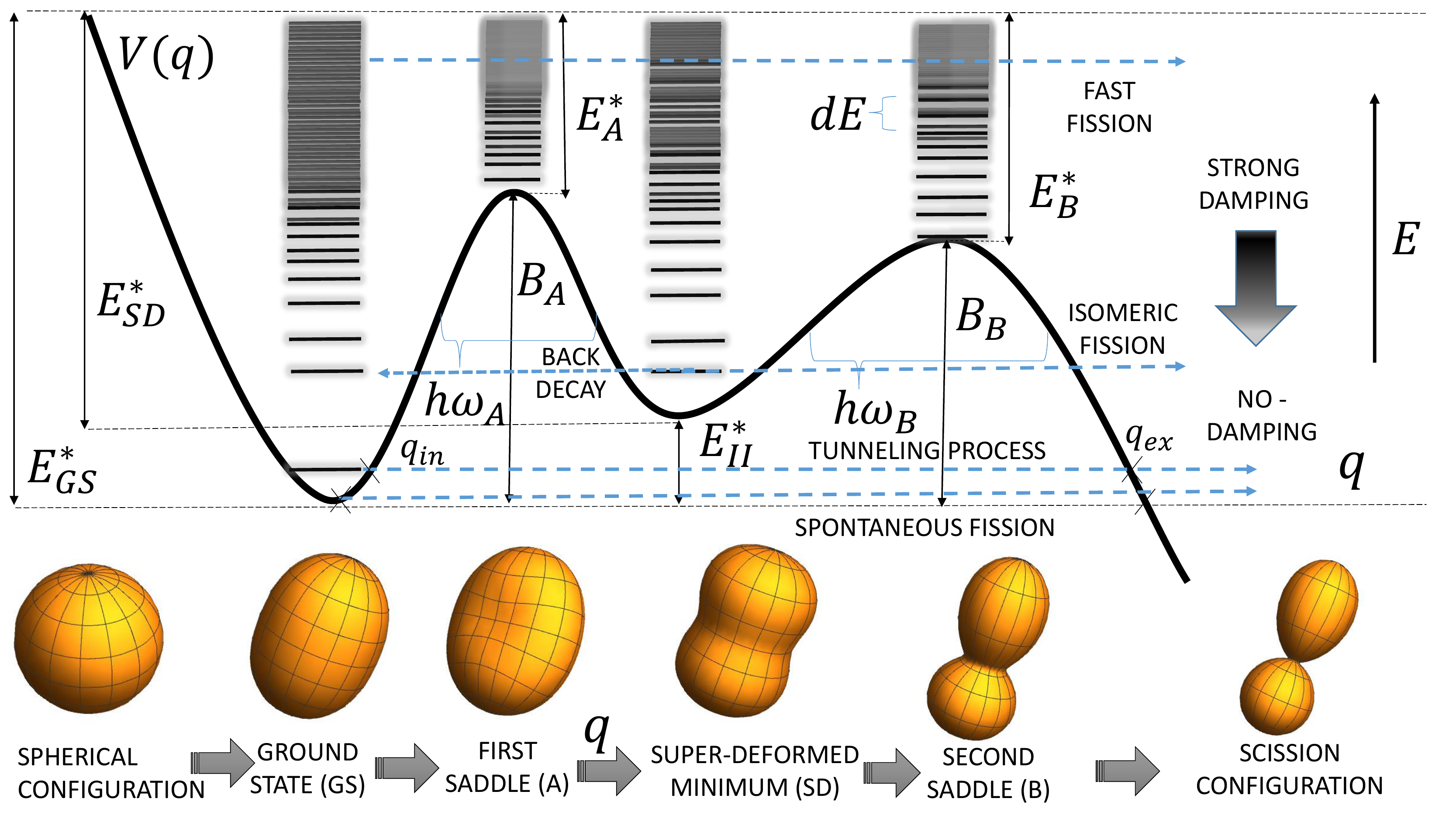}}
\vspace{-0mm}
\caption{{\protect Schematic one-dimensional picture of a double-humped
  fission barrier typical of actinides along the effective deformation
 parameter $q$.
  The index $A(B)$ stands for the first (second) saddle while $GS(SD)$ for the first (second) minimum.
  The fission barrier heights $B_{A(B)}$ and curvatures $\hbar \omega _{A(B)}$
 are indicated.
 Densities of states and nuclear shapes corresponding to minima, saddles and a possible shape at the scission point
 are depicted below.}}
\label{schem}
\end{figure}


 Although the fission barrier heights $B$ are not directly measurable
 quantities, i.e. they are not quantum observables, they are very useful
 in estimating nuclear fission rates. As the activation energy $E_a$ (per mole)
 in chemistry gives a rate $k$ of a chemical reaction at temperature $T$ via
 the Arrhenius law: $\Gamma =A e^{-E_a/RT}$ ($R$ - the gas constant; $A$ - the
 frequency factor) \cite{tHoff,Arrhe}, the fission barrier gives the fission
 rate $\Gamma ^{fiss}$ of an excited (as they usually are in nuclear reactions)
 nucleus via:
 $\Gamma ^{fiss} \sim e^{-B/T_{eff}}$, where $T_{eff}$ is an effective
  temperature (in energy units) derived from the excitation energy.

 Let us give a few examples of nuclear reactions, to the analysis of which enters the fission barrier -
 that is, the "activation" threshold for the fragmentation of the nucleus into two fragments:

\begin{itemize}
\item the competition between neutron emission from, and fission of, a
 compound nucleus in fusion-fission reactions which determines the survival
 probability of the evaporation residue i.e. the probability that a
 superheavy nucleus will be created. This probability ratio is
 extremely sensitive to changes in $B$;
\item the growth of the total cross sections for the formation of the superheavy nuclei around Z=118;
\item the competition between fission and other decay modes in the neutron capture reactions on neutron-rich nuclei
 lying on the path of the so-called r-process of the nucleosynthesis;
\item fission barriers enter to the codes estimating induced fission
cross-section as a function of energy, relevant to the reactor physics;
\item spontaneous fission in which not only $B$ alone, but the whole energy
 landscape, together with the mass parameters, determine fission half-lives.
\end{itemize}
 A non-observable status of the fission barrier, again in analogy to that of
 the activation energy in chemistry, is reflected in its possible dependence
 on the reaction type and the excitation energy (effective temperature) range.
 This leads to some uncertainty in the calculations of fission barriers.

  One can imagine an excited nucleus as a set of nucleons bouncing from
   walls of the slowly changing selfconsistent mean potential. The
  motion of the walls results from averaged motions of all
  nucleons and is 1-3 orders of magnitude slower than them. Separating
  slow, collective degrees of freedom of the potential walls from the
  fast ones, derived from motions of single nucleons, was never exactly
  accomplished in nuclear physics. Nevertheless, it is used in various
  simplified models to describe reactions like fission.

  Within the mean-field theory, the wave function of a nucleus is assumed
 as a product state of individual nucleons occupying orbitals of the
 selfconsistent potential which depends on collective coordinates, denoted $q$.
 Take as $q$, for example, the elongation of a nucleus.
 Energies of s.p. states as functions of $q$ have many avoided
 crossings between the g.s. and scission.
 Occupations of s.p. orbitals which at each $q$ correspond to the lowest
 energy of a nucleus are called adiabatic, and in the HF it means that
 nucleons occupy the lowest orbitals. For a nucleus elongating with
 a velocity ${\dot q}>0$ some nucleons may not follow adiabatic
 occupations at level crossings. If a nucleon occupies the same upsloping level
 through the crossing one says that its behaviour is diabatic.
 The uncertainty as to the adiabatic or diabatic character of crossings
 in fission directly translates into an uncertainty in the fission barrier
 height $B$. This problem concerns also activation energy
 in chemical reactions.

 For an idealized isolated crossing of two levels with
 constant and opposite slopes $d\varepsilon/dq$, the interaction $V$, and at
 constant velocity ${\dot q}$, there is an exact solution
 for the probability that the particle occupying at time
 $t\rightarrow -\infty$ the lower adiabatic level
 $E_1=\frac{1}{2}[(\varepsilon_1+\varepsilon_2)-\sqrt{(\varepsilon_2-
\varepsilon_1)^2+4V^2}]$ will occupy the upper adiabatic level
 $E_2=\frac{1}{2}[(\varepsilon_1+\varepsilon_2)+\sqrt{(\varepsilon_2-
 \varepsilon_1)^2+4V^2}]$ at $t\rightarrow \infty$ \cite{LZS}:
\begin{equation}
 P = \exp\left(-\frac{2\pi}{\hbar}\frac{V^2}
 {{\dot q}|\frac{d}{dq}(\varepsilon_2-\varepsilon_1)|}\right)  .
\end{equation}
 One should notice that the jump to the upper adiabatic state means that
 the crossing is {\it diabatic} - the particle will end in the
 state with the same slope.
 Generalizing this particular case one might say that a large interaction $V$,
 a small velocity ${\dot q}$ and small relative slope
 $|d/dq(\varepsilon_2-\varepsilon_1)|$
 help to keep adiabatic configuration, while the opposite conditions
 enhance diabatic continuation \cite{LZS,Hill1953,Schutte1978,Nazarewicz1993}.
 However, one should realize that the real processes last for a finite time,
  level slopes and the velocity change and consecutive crossings can interfere.
\begin{figure}[htp]
 \centering
\includegraphics[width=14cm,height=8cm,angle=0.0]{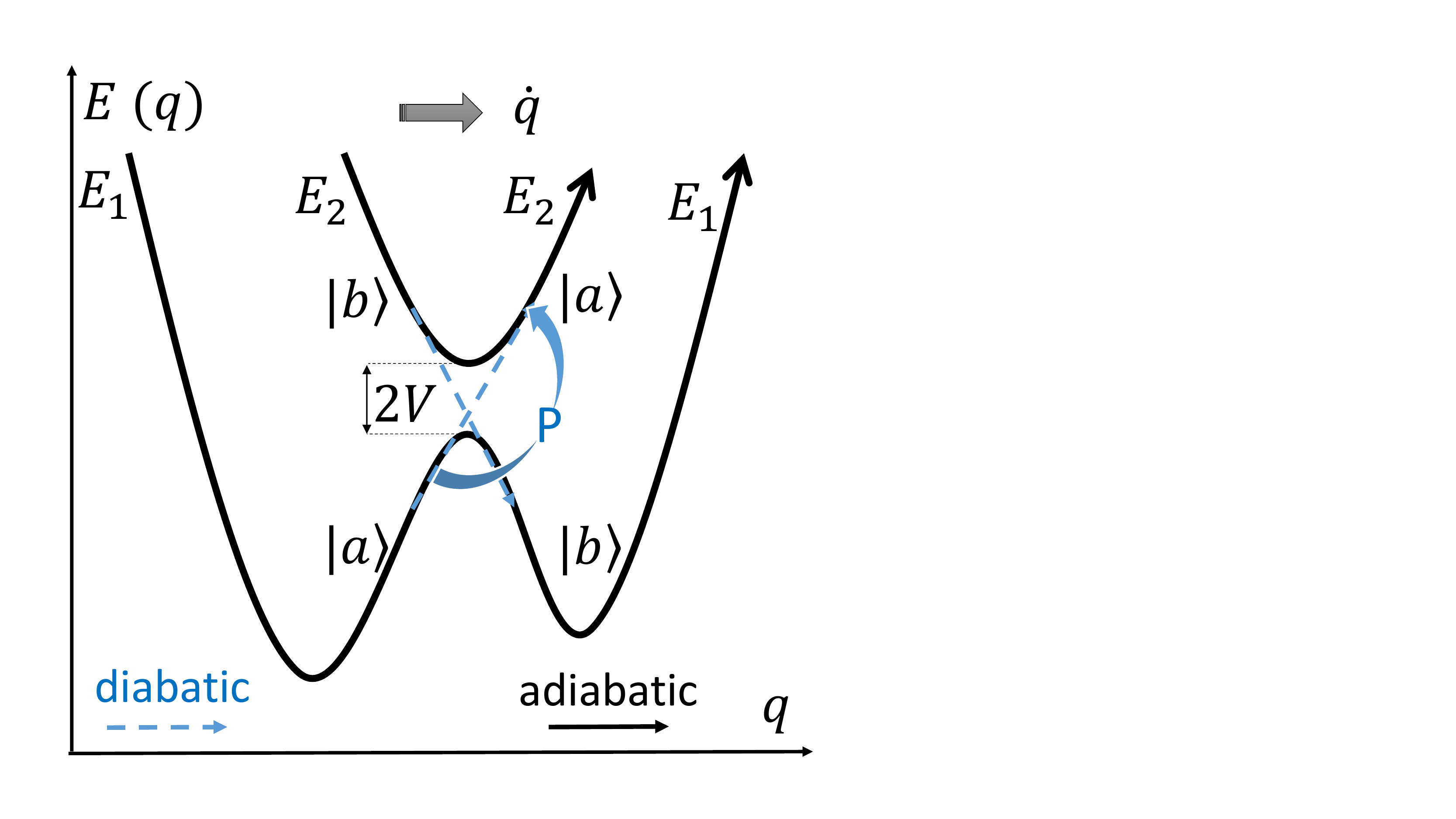}
\caption{\label{LZS}
 Diabatic (blue arrow and dashed lines) vs. adiabatic (black lines) transition
 at the crossing of adjacent adiabatic levels $E_{1}$ and $E_{2}$ with the
 minimal distance $2 V$.}
\end{figure}
 Adiabatic vs diabatic scenario is schematically
  illustrated in Fig. \ref{LZS} for two adiabatic energy levels $E_1$ and $E_2$
  which exchange quantum characteristics of nonintracting states
 $\mid a \rangle$ and $\mid b \rangle$ in the region of crossing.
 The diabatic tendency to preserve level characteristics, including quantum
 numbers, is indicated by the blue arrow, and the adiabatic one, with the
 residual interaction leading to their exchange, by the black lines.
   A difference between the diabatic and adiabatic scenario
 is especially relevant for odd-$A$ and
 odd-odd nuclei. It seems that for them, the sharp crossings of levels
 occupied by the odd particle exclude the strictly adiabatic scenario.
 It is known that the diabatic effect of conserving the $K$ quantum
 number (projection of the angular momentum of a nucleus on its symmetry axis)
  leads to a large increase in the calculated fission barriers,
 see, e.g., \cite{Kisomers}.
Unfortunately, there is no sufficiently reliable theory that would predict
 the importance of diabatic effects.



 For the sake of further discussion we will distinguish three regimes
 according to the excitation energy of a nucleus.
  In particular, the induced fission with $E^*_{GS}>>B_{A(B)}$
 should be separated from the spontaneous one occurring from the ground state.
Only in these extreme cases, when the energy is either much higher than the
 barrier height (then the details of the barrier do not actually matter much)
and when we are dealing with the quantum tunneling deep under the barrier,
 one can treat fission as a one-step process
The region of the excitation energy just above the first or second saddle or
 not much smaller than them seems the most difficult to describe.
 There is no fundamental theory for the fission process in this energy regime.
 On the other hand, it is crucial from the point of view of
 applications e.g. in the reactor physics. Different mechanisms of fission
 depending on the energy available to the system are illustrated in
 Fig. \ref{schem} .
 The fission barrier $B_{A(B)}$ helps in the understanding of each of them.

\subsection {Regime: $E_{GS}^{*} \gg B_{A(B)}$ - statistical limit}

In 1938 E.~Wigner introduced Transition State Formalism (TSF) to calculate
 chemical reaction rates \cite{Wigner}.
Later Bohr and Wheeler \cite{Bohr21939} realized that this formalism may be used to estimate
 the fission probability and that this process is mainly governed by the number
 of states above the fission barrier.

Due to differences in the shape of the nucleus,
the energy levels above the fission barrier and above the ground state
 are distributed differently.
 For the excitation energy well above the {\it highest saddle point}
the  TSF predicts the fission decay width by integrating the density of states:$ \rho_{A(B)}(E^{*}_{A(B)} - E)$
in the energy window: $E^{*}_{A(B)} = E^{*}_{GS}-B_{A(B)}$ (see Fig. \ref{schem}):
\begin{equation}
 \Gamma ^{fiss} _{A(B)} (E^{*}_{GS}) = \frac {1}{2\pi \rho (E^{*}_{GS})}  \int_{0}^{E^{*}_{A(B)}} \rho_{A(B)}(E^{*}_{A(B)}-E) dE  .
\label{Wigner}
\end{equation}
 Here, $\rho(E)$ denotes the density of states of a nucleus at given
 deformation as a function of energy.
The fission rate is controlled mainly by the height of the fission barrier
 $B_{A(B)}$ as it determines the upper limit of the integration in
 Eq. (\ref{Wigner}), that is, the energy window containing the
 appropriate number of nuclear states.

In the microcanonical ensemble the logarithm of the phase-space volume of the
excited nucleus defines the entropy \cite{Landau}, while its derivative with
respect to energy gives the level density (for simplicity here:
 $\rho_{A(B)}(E^{*}_{A(B)}-E) = \rho $):
\begin{equation}
\rho = e^{S} \frac{dS}{dE^*} \Rightarrow \overrightarrow{T =
 \frac{dE*}{dS}|_{V=const}} \Rightarrow \rho= \frac{e^S}{T}  .
\label{simplerho}
\end{equation}
We obtain this relationship by using the Maxwell relation for the
 thermodynamic potentials assuming the volume conservation during the process
(in nuclear physics temperature is measured in energy units).
Using the Fermi gas approximation (with the constant level density parameter
 $a$) which
 is reasonable for the excitation energy high above the barrier one has:
 $S=2aT$ and $E^*=aT^2$.
Inserting these into equation Eq.  (\ref{Wigner}) and (\ref{simplerho}) we
 obtain (with the accuracy of the integration constant)
 the following expression for the fission rate:
\begin{equation}
  \Gamma ^{fiss} _{A(B)} \sim \frac{e^{S_{A(B)}}-1}{e^{S_{GS}}}
 .
\end{equation}
Since $S_{A(B)} >> 1$,
 we have:
\begin{equation}
 \Gamma ^{fiss} _{A(B)} \sim e^{2\sqrt{aE_{A(B)}^*}-2\sqrt{a E_{GS}^*}} \Longrightarrow  \Gamma ^{fiss} _{A(B)} \sim e^{2\sqrt{a(E_{GS}^*-B_{A(B)})}-2\sqrt{a E_{GS}^*}}.
\end{equation}
For $E_{GS}^*\gg B_{A(B)}$ one can write:
\begin{equation}
\sqrt{a(E_{GS}^*-B_{A(B)})}-\sqrt{aE_{GS}^*} \approx \frac{-B_{A(B)}}{T}.
\end{equation}
This led us to the Arrhenius rate-formula:
\begin{equation}
  \Gamma ^{fiss} _{A(B)} \sim \exp\left(-\frac{B_{A(B)}}{T}\right) .
\label{Arh}
\end{equation}
For low excitation energies, the assumption of an independent movement of
 nucleons is untenable, as the density of low-lying nuclear energy levels
 shows much structure.

 The Eq. \ref{Arh} is only valid for the angular momentum zero i.e. for
 the non-rotating nuclear system.
For a non-zero angular momentum, assuming the axially symmetric shape
  at the saddle point, the rotational energy for the total angular momentum
 $J$ and its projection onto symmetry axis $K$ takes the form:
\begin{equation}
  E_{rot} (J,K) = \frac{\hbar^2 (J^2-K^2)}{2J_{\perp}} - \frac{\hbar^2 K^2}{2J_{\parallel}} .
\label{Erot}
\end{equation}
The saddle moments of inertia: perpendicular - $J_{\perp}$ and parallel -
 $J_{\parallel}$ to the symmetry axis, may be calculated
from the rigid body model or treated as the empirical fit parameters.
In a first approximation, one can simply add the rotational energy
 Eq. (\ref{Erot}) at the ground state and at the saddle
 (different due to different deformations).
 A more refined recipe would be to add the rotational energy $E_{rot}(J, K)$
 at each deformation and find the fission barrier depending on the angular
 momentum. Generally, as the moments of inertia increase with deformation,
 fission barrier decreases with J.

For high excitation energies, the phenomenon of dissipation begins to play an important role.
 H.~Kramers was one of the first \cite {Kramers} to consider such dynamical
 corrections.
Assuming the one-dimensional fission path \cite{Kramers},
a parabolical shape for the ground state minimum (of curvature $\omega_{GS}$),
and an inverted parabola shape of the top of the barrier
 (with curvature - {$\omega_{A(B)}$}),
one obtains, in the limit of large dissipation:
\begin{equation}
\Gamma ^{fiss} _{diss}  \simeq  \frac{\omega_{GS}}{\omega_{A(B)}}
( \sqrt{\frac{\gamma^2}{4} + \omega_{A(B)}^2} - \frac{\gamma}{2})
e^\frac{-B_{A(B)}}{T}
 \Rightarrow \overrightarrow{\gamma \gg \omega_{A(B)}} \Rightarrow
\Gamma ^{fiss}_{overdamped} \simeq \frac{\omega_{A(B)}
\omega_{GS}}{\gamma} e^\frac{-B_{A(B)}}{T} .
\end{equation}

The condition ($\gamma \gg \omega_{A(B)}$) is called the overdamped limit.
The friction coefficient  $\gamma$ controls the intensity
of the stochastic forces $\eta(t)$  responsible for the interaction of the
collective motion with the thermal bath of other degrees of freedom (the environment) via the autocorrelation function
$\langle \eta(t) \eta(t') \rangle \approx  \gamma T \delta (t -t')$. This stochastic force (based on rather heuristic arguments)
is added to the conservative driving force $-\frac{d V(q)}{dq}$, giving a
 stochastic dynamic description of the fissioning system.
At the limit of an overdamped stochastic process, the corresponding classical Langevin equation of motion has the form:

\begin{equation}
\gamma \frac{dq}{dt} = - \frac{d V(q)}{dq} + \eta(t) .
\end{equation}

The dissipative dynamics approach has proved very successful in modeling the fission process.
There are a number of generalizations of this method due to many dimensions, different friction regimes or types of the stochastic noise.
 A broad overview can be found in \cite{Feldmeier,Hanggi1990,Talkner,Frobrich}.

\subsection {Regime: $E_{GS}^{*} \ll B_{A(B)}$ - quantum tunneling limit}

The collective tunneling limit is usually treated in the quasiclassical approximation which leads to a WKB-type formula \cite{LYN}.
The fission rate up to subexponential corrections is given by:
\begin{equation}
\Gamma  \approx P^{WKB} \approx e^{-2S(L)},
\end{equation}
where $S(L)$ is action integral:
\begin{equation}
S(L) = \int_{q_{in}}^{q_{ex}} \sqrt { \frac{2}{\hbar^{2}} \{ M_{L}(q)
[ V_{L}(q) - E_{0}]\}} dq ,
\end{equation}
calculated along the optimal fission path $L$ in a multi-deformation space,
for which $S(L)$ is the smallest.
The application of the variational principle makes it possible to find the optimal path.
Here, $V_{L}(q)$ is the potential energy,  $M_{L}(q)$ is the effective
inertia along the trajectory $L$. $E_{0}$ is the initial energy of a
fissioning nucleus. The parameter $q$ specifies the position of
a point on the trajectory $L$, with $q_{in}$ and $q_{ex}$ corresponding to
the entrance and exit points of the barrier, i.e., to the classical
turning points determined by $V_{L}(q) = E_{0}$.
These points are schematically shown in Fig.1.

The effective inertia $M_{L}(q)$ associated with the fission motion along the trajectory $L$
is usually calculated in the frame of the cranking approximation \cite{Sobiczewski69}.
Because it goes beyond the scope of this chapter, we refer the reader to the book \cite{Pomorski}.





\subsection {Regime: $E_{GS}^{*} \simeq B_{A(B)}$ - damping limit}

One can try to understand physics in this limit or even calculate the
 corresponding effective tunneling coefficients based on the idea of the
 optical
 potential \cite{LYN,Sin,Sin2,Goriely2009}.
 The transmission through a multi-humped fission barrier depends on the degree of damping of the vibrational states in
the potential minima. In the frame of this concept, the strong damping limit,
 marked in the Fig. \ref{schem} for energies
close to $B_{A}$ and $B_{B}$, means that the second minimum begins to absorb
 particles passing through it.
 A fraction of the particle stream not absorbed by the second minimum may
 pass further through the second barrier, usually different from the first
 one in its shape and height.
These differences cause different probabilities of transmission over each of
 them.
In the strong damping regime, assuming full absorbtion by the second minimum,
one can write the average transmission ratio as:
\begin{equation}
\langle T \rangle  \simeq \left[ \frac{1}{T_{A}}+\frac{1}{T_{B}} \right]^{-1}
\Longrightarrow \langle T \rangle  \simeq  \frac{T_{A}T_{B}}{T_{A} + T_{B}}.
\label{Tdamp}
\end{equation}
In this approximation, coefficients of transmission through successive barriers
 are independent of each other.

In practical calculations,
the individual character of low-lying excited states should be taken into account.
The discrete transition states (denoted here by the index d ) are the
 rotational levels built on vibrational or non-collective band heads,
characterized by a given set of quantum numbers (angular momentum J,
 parity $\pi$, and angular momentum projection on the nuclear symmetry axis K).
We may assume that the fission process occurs through such analog states, i.e., retaining the quantum characteristics of the compound nucleus states.
These so-called transition states are located on top of the appropriate
 saddle point, and due to their discrete nature,
they should be separated from the continuous spectrum by assigning an individual probability of decay to each of them.
In general, each such state has a specific fission barrier $B^{d}_{A(B)}$ and,
 consequently, the probability of penetration thereof.
 For nucleus energy above the barrier, $E_{GS}^* \geq B^{d}_{A(B)}$,
 one uses the Hill-Wheeler barrier transmission coefficient \cite{Hill1953,LYN}:
\begin{equation}
T^{HL}_{A(B)}(E) = \frac{1}{1+exp[\frac{2\pi (B^{d}{A(B)}-E) }{\hbar \omega_{A(B)}}]}.
\end{equation}
If available energy is below the barrier, $E_{GS}^* < B^{d}_{A(B)}$,
 then we can use the previously introduced tunneling probability $P^{WKB}$.
Transmission coefficients  $T_{A(B)}$, which one should then use in
 Eq. (\ref{Tdamp}) are written in the following form:
\begin{equation}
 T_{A(B)} = \sum_{d} P^{HL(WKB)}_{A(B)} (E) + \int_{E_{d}}^{E^{*}_{d A(B)}} P^{HL(WKB)}_{A(B)}
 (E) \rho_{A(B)}(E^{* }_{d A(B)}-E) dE.
\end{equation}
Note that this time the integration takes place from a specific energy of the discrete state above the saddle ($E_{d}$) in the energy window defined by the
 diabatic barrier $B^{d}_{A(B)}$, characteristic for it, $E^{*}_{d A(B)} =
 E_{GS}^{*}- B^{d}_{A(B)}$.
In this way we have extracted (following the authors \cite{Goriely2009})
a discrete part of the spectrum above the saddle.
 After normalizing to the density of the available states over the ground
 state, one obtains the average effective fission width in the over-damped
 regime:
\begin{equation}
\langle  \Gamma ^{fiss} \rangle = \frac{\langle T \rangle}{ 2 \pi  \rho (E^{*}_{GS})}.
\label{meanG}
\end{equation}
This is a heuristic and approximate, but pragmatic approach as it allows to calculate the probabilities of induced fission.
Unfortunately, there is still no fundamental theory that would allow the reconstruction of cross-sections in this process,
that are key for applications.
The main reason for this state of affairs is the fact that we still do not know how to correctly calculate the fundamental quantity occurring in most of the integrals presented, namely the total density of nuclear states located both above the ground state and above the appropriate saddle.
There are only a number of approximate methods to do this \cite{hilary,hilary2,Blin,Aberg1,Aberg2,Aberg3,Aberg4,pomorska1,Senkov,Alhassid,Rah2021}.

\section{Macroscopic - Microscopic method}

 The macroscopic-microscopic method is an approximation to the full
 selfconsistent Hartree-Fock(-Bogolyubov) theory. It is based on the
 observation that differences between nuclear binding (or masses) and their
 liquid-drop model estimates
 oscillate with $Z$ and $N$ with the amplitude $\le 10$ MeV.
 The minima of these oscillations correspond to particularly stable
 magic nuclei. This difference, termed the shell correction,
 has its origin in the bunching of quantum s.p. levels i.e. the non-uniformity
 of the density of quantum s.p. states per energy interval.
 This genuinely quantum
 phenomenon occurs in many spectra and in the case of atomic nuclei leads to
 shells containing $\approx A^{2/3}$ states, with the energy distance
 between them of $\hbar \Omega\approx$5 - 10 MeV.
  V.~Strutinsky proposed a simple method to calculate shell correction \cite{Strutinsky1969,Strutinsky21969}
 based on the idea that
 a difference between the real binding energy and its value given by the
 liquid-drop formula could
 be expressed as a difference between the sums of single-particle energies in
 a phenomenological s.p. potential: one for the
  quantum one-body density and the other for the density averaged
 over shells.
 The averaging should eliminate shell non-uniformities, thus its energy range
 should be of the order of $\hbar\Omega$. The averaged density of s.p.
 states should be close to the quasiclassical one, without shell oscillations,
 but reflecting a smooth increase in energy $\sim \sqrt{E-V_0}$ (with $V_0$ -
 the bottom of the potential well), characteristic of three-dimensional systems.
 Below, following M.~Brack et al. 1972 \cite{Brack1972}, we present the shell correction method
 in connection to the mean-field theory.

  \subsection{Shell Correction vs the Mean Field}

 Energy of a nucleus in a state represented by a Slater determinant built out
 of single-nucleon states $\mid\phi_k\rangle$,
  with the Hamiltonian consisting of the kinetic energy
 ${\hat t}$ and a two-body nucleon-nucleon interaction ${\hat {\rm v}}$,
 ${\hat H}={\hat t}+{\hat {\rm v}}$, can be expressed as a
 function of the one-body density ${\rho}$ which,
 when understood as an operator, can be written as:
\begin{equation}
 {\hat \rho}=\sum_{n\ occ} \mid\phi_n\rangle\langle\phi_n\mid ,
\end{equation}
 with $n$ - indices of the occupied orbitals. (The one-body density should not
 be confused with the energy density of states of the whole nucleus,
 customarily denoted by the same symbol in the preceding section.)
 In any orthogonal s.p. basis, we have:
\begin{equation}
 E(\{\phi_k\})=E(\rho)=\sum_{\mu \nu} t_{\mu \nu}\rho_{\nu \mu}+
 \frac{1}{2}\sum_{\mu \nu \gamma \delta}(v_{\mu \nu \gamma \delta}
 -v_{\mu \nu \delta \gamma})\rho_{\delta \nu}\rho_{\gamma \mu} ,
\end{equation}
 with $v_{\mu \nu \gamma \delta}$ - matrix elements of the interaction.
 The functional derivative of energy $E(\{\phi_k\})$ with respect to the
 complex-conjugate function $\phi_m^*$ equals:
 \begin{equation}
 \delta E(\{\phi_k\})/\delta \phi_m^*=
 {\hat h}(\rho)\mid\phi_m\rangle ,
\end{equation}
 where ${\hat h}(\rho)={\hat t}+{\hat V}$
  is the one-body Hamiltonian with the mean-field potential dependent on the
  density $\rho$, i.e. on the occupied orbitals:
 \begin{equation}
 \label{poten}
   V_{\mu \nu}(\rho)=\sum_{\gamma \delta} (v_{\mu \gamma \nu \delta}-
  v_{\mu \gamma \delta \nu})\rho_{\delta \gamma} .
\end{equation}
 One-body density can be presented as a function: $\rho({\bf r},{\bf r}')=
 \langle{\bf r}\mid {\hat \rho}\mid{\bf r}'\rangle=\sum_n \phi_n({\bf r})
  \phi^{\dagger}_n({\bf r}')$. If some function
  $\rho({\bf r},{\bf r}')$ is a selfconsistent one-body density solution to
 the Hartree-Fock problem, there is a basis of s.p. states
 $\mid\phi_\mu\rangle$, a subset $\mid\phi_n\rangle$ of which forms the density
  $\rho$, that fulfil HF equations:
 \begin{equation}
  {\hat h}(\rho)\phi_{\mu} = ({\hat t}+{\hat V})\phi_{\mu} =
  e_{\mu} \phi_{\mu} ,
 \end{equation}
 with $e_{\mu}$ the HF s.p. energies. One can construct such solutions
 for various prescribed values of multipole moments which would correspond to
 various nuclear deformations.

 A functional dependence of energy $E(\rho)$ on $\rho({\bf r},{\bf r}')$ can
 be considered generally even for functions which {\it do not represent
 any Slater determinant}. One can formally define the functional derivative
  $\delta E(\rho)/\delta\rho({\bf r},{\bf r}')$ as the matrix element of some
 one-body Hamiltonian $\langle{\bf r}'\mid {\hat h}(\rho)\mid{\bf r}\rangle$.
 With such a more general definition, a difference in energies
 calculated for such generalized densities, one $\rho^{(1)}$ and the other
  $\rho^{(2)}$, reads:
 \begin{equation}
 \label{HF}
  E(\rho^{(1)})-E(\rho^{(2)})=
 {\rm Tr}\ {\hat h}(\rho^{(2)})(\rho^{(1)}-\rho^{(2)})\ +
 \ {\rm terms} \sim (\rho^{(1)}-\rho^{(2)})^2  ,
 \end{equation}
 where, in any orthogonal basis, ${\rm Tr}\ A B = \sum_{\mu \nu} A_{\mu \nu}
 B_{\nu \mu}$.

  Now we consider a density ${\tilde \rho}$, obtained from $\rho$ by a
 procedure of averaging over the shell structure, the
 details of which will be specified later. In a broad sense, one can think of
 ${\tilde \rho}(Z_0,N_0)$ as obtained by averaging one-body HF densities
 over a range
 of neutron and proton numbers $N$ and  $Z$ around $N_0$ and $Z_0$.
  It is assumed that the density
 ${\tilde \rho}$ is close
 to $\rho$, and the mean-field ${\hat h}({\tilde \rho})$ is close to
 ${\hat h}(\rho)$, so that $\delta\rho=\rho-{\tilde \rho}$ and
 the difference in s.p. potentials $\delta V =
  {\hat h}(\rho)-{\hat h}({\tilde \rho})=V-{\tilde V}\sim \delta\rho$
 are considered as small of the first order. In general, the density
 ${\tilde \rho}$ does not correspond to any Slater determinant so there are no
 orbitals building it and no HF equations which they would fulfil.
 Nevertheless, there is a basis of s.p. wave functions defined by the
 eigenequation:
 \begin{equation}
  {\hat h}({\tilde \rho})\psi_{\mu} = ({\hat t}+{\hat {\tilde V}})\psi_{\mu} =
 \epsilon_{\mu} \psi_{\mu} .
 \end{equation}
 We interpret ${\hat h}({\tilde \rho})$ as the {\it phenomenological s.p.
 Hamiltonian} and its eigenfunctions $\psi_{\mu}$ as the corresponding
 {\it phenomenological s.p. wave functions}. Those of them which correspond to
 occupied selfconsistent states $\phi_n$ define the {\it phenomenological
 s.p. density}, $\rho^{S}=\sum_{n\ occ} \mid\psi_n\rangle\langle\psi_n\mid$. As
 $h(\rho)=h({\tilde \rho})+\delta V$, the difference
 between $\phi_{\nu}$ and $\psi_{\nu}$ follows from the perturbation
 expansion:
 \begin{equation}
  \mid\phi_{\nu}\rangle=\mid\psi_{\nu}\rangle+\sum_{\mu\ne\nu}
 \mid\psi_{\mu}\rangle
 \frac{\langle\psi_{\mu}\mid \delta V\mid\psi_{\nu}\rangle}
 {\epsilon_{\nu}-\epsilon_{\mu}} + .....
 \end{equation}
 Since to the first order in $\delta V$: $\rho-\rho^{S}=
 \sum_{n\ occ} \mid\delta\psi_n\rangle\langle\psi_n\mid+
  \mid \psi_n\rangle\langle\delta\psi_n\mid$, where $\delta\psi_n=
 \phi_n-\psi_n$, and $\langle \delta\psi_k\mid\psi_k\rangle=0$, one obtains
 that ${\rm Tr}\ {\hat h}({\tilde \rho})(\rho-\rho_S)$ is of the
 second order in $\delta V$.

 As the difference between $\rho$ and $\rho^S$ is in the first order
 proportional to $\delta V$, the difference between averaged densities,
 ${\tilde \rho}$ and ${\tilde \rho}^S$ would be proportional to the
 averaged $\delta V$. From the formula (\ref{poten}) for $V_{\mu \nu}(\rho)$
 this is proportional to the averaged $\delta \rho$ and that is zero.
 Therefore, it is natural to expect that the difference
 ${\tilde \rho}-{\tilde \rho}^S$ is of the second order in $\delta \rho$.

 Then, as $\delta \rho= (\rho-\rho^S)+(\rho^S-{\tilde \rho}^S)+
 ({\tilde \rho}^S-{\tilde \rho})$, it follows from the above that
 the difference between ${\rm Tr}\ {\hat h}({\tilde \rho}) \delta\rho$ and
 ${\rm Tr}\ {\hat h}({\tilde \rho}) (\rho^S-{\tilde \rho}^S)$
  is of the second order in $\delta \rho$.
  Hence, from Eq. (\ref{HF}), subsituting $\rho$ for $\rho^{(1)}$ and
 ${\tilde \rho}$ for $\rho^{(2)}$, one obtains:
 \begin{equation}
  E_{HF}(\rho)-E({\tilde \rho})=
 {\rm Tr}\ {\hat h}({\tilde \rho})(\rho^S-{\tilde \rho}^S)\ +\ {\rm terms}
  \sim (\delta \rho)^2  .
 \end{equation}
 This equation, sometimes called the Strutinsky energy theorem, is the basis
 of the shell correction method. It states that the selfconsistent HF energy
 may be presented as a sum of two terms: $E({\tilde \rho})$ which
 smoothly depends on $Z$ and $N$ and ${\rm Tr\ }{\hat h}
 (\rho^S-{\tilde \rho}^S)$ which exclusively derives from the
 {\it phenomenological}, deformation-dependent s.p. Hamiltonian and contains
 all the first order effects of the shell structure. More specifically,
 the first order effects of the shell structure are contained in the sum
 of s.p. energies of occupied phenomenological levels ${\rm Tr}\ {\hat h}
 \rho^S$, as those in ${\rm Tr}\ {\hat h}{\tilde \rho}^S$ are at most of
 the second order in $\delta \rho$.

 This lends argument to the following reasoning:
  If we replace the smoothed HF energy $E({\tilde \rho})$ by a
 liquid-drop energy $E_{LD}$, and assume
 that a phenomenological single-particle Hamiltonian ${\hat h}$ has the same
 spectrum around the Fermi energy as ${\hat h}({\tilde \rho})$, we obtain
 to the first order in $\delta \rho$:
 \begin{equation}
  E_{HF}(\rho)\approx E_{LD}+ {\rm Tr}\ {\hat h}(\rho^S-{\tilde \rho}^S) =
             E_{LD}+\delta E.
 \end{equation}
 If we further adjust the parameters of the liquid-drop energy so that
 $E_{LD}+\delta E$ with a given phenomenological potential fit
 the experimental binding energies we may hope to incorporate the second order
 terms of the shell correction into $E_{LD}$ and improve the overall
 description of data beyond the first order agreement of the Strutinsky
 energy theorem.

  \subsection{Calculation of the Shell Correction }

 To calculate the shell correction one uses the density of single-particle
 states of ${\hat h}$ per energy interval which is just a sum of delta
 functions at single-particle energies,
 $g(\epsilon)=\sum_{\nu}\delta(\epsilon-\epsilon_{\nu})$,
 and its smoothed version (defined below) ${\tilde g}(\epsilon)$:
 \begin{equation}
       \delta E =  \int \epsilon (g(\epsilon)-{\tilde g}(\epsilon)) d\epsilon
 = U-{\tilde U} ,
 \label{shcor}
 \end{equation}
 with:
 \begin{equation}
  U = \sum_{\nu}  n_{\nu} \epsilon_{\nu} ,
 \end{equation}
  where $n_{\nu}$ are occupation numbers (either 0 or 1), and
 \begin{equation}
 {\tilde U} =  \int_{-\infty}^{\tilde \lambda}\epsilon
 {\tilde g}(\epsilon)d\epsilon.
 \end{equation}
  The quantity ${\tilde \lambda}$ is the Fermi energy for the smoothed density,
 determined by the fixed number of particles.
 The whole procedure is made separately for neutrons
 and protons, and the shell correction is the sum of their separate
 contributions, $\delta E=\delta E_n+\delta E_p$. Hence the equation fixing
 ${\tilde \lambda}_n$ is:
 \begin{equation}
\label{Strutlam}
 N = \int_{-\infty}^{{\tilde \lambda}_n} {\tilde g}_n(\epsilon)d\epsilon,
 \end{equation}
 and the analogous one fixes ${\tilde \lambda}_p(Z)$.
  Usually one assumes the Kramers degeneracy of s.p. levels and then one
 has sums over half of s.p. levels multiplied by 2.

The "smooth" density is not unique, but arguments can be made for
choosing it in a form of convolution (folding) of the density $g(\epsilon)$
 with some smoothing function $\xi$ with a width ${\bar \gamma}> \hbar\Omega$,
\begin{equation}\label{smooth_den_def}
{\tilde g}(\epsilon)= \frac{1}{\bar \gamma}
\int^{\infty}_{-\infty} g(\epsilon')
 \xi\left(\frac{\epsilon-\epsilon'}{\bar \gamma}\right)
d \varepsilon^{\prime} = \frac{1}{\bar \gamma}\sum_{\nu}
 \xi\left(\frac{\epsilon-\epsilon_{\nu}}{\bar \gamma}\right) .
\end{equation}
 It follows from the finite width ${\bar \gamma}$ and the condition
 (\ref{Strutlam}) that the shell correction is determined exclusively
 by the s.p. levels lying, roughly, not much farther than ${\bar \gamma}$
 from the Fermi level.

 Most often one takes for $\xi$ a folding function $f_p$ of the Gaussian type,
 being the formal expansion of the $\delta$-function,
truncated to the first $p$ terms (with $p$ even):
\begin{equation}\label{delta_expand_approx}
f_p(x) = \frac{1}{\sqrt{\pi}} \sum_{n=0}^{p} C_n H_n(x)
e^{-x^2},
\end{equation}
with
\begin{equation}\label{C_n}
C_n= \frac{1}{2^n n!} H_n(0) = \left\{
\begin{array}{ll}
\frac{(-1)^\frac{n}{2}}{2^n(\frac{n}{2})!} & \textrm{for even $n$} \\
0 &  \textrm{for odd $n$.} \\
\end{array} \right .
\end{equation}
Then, one obtains the averaged density:
\begin{equation}
 {\tilde g}(\epsilon)=\frac{1}{\bar \gamma\sqrt{\pi}}\sum_{\nu} e^{-u_\nu^2}
\sum_{n=0}^{p}C_n H_n(u_\nu),
\end{equation}
with $u_\nu=(\epsilon-\epsilon_\nu)/{\bar \gamma}$.

The shell correction (\ref{shcor}) is strongly correlated with the local density of states around the
Fermi energy $\epsilon_F$; $\delta E > 0$
if there are more states there than it would follow from ${\tilde g}(\epsilon_F)$, $\delta E < 0$ if there are less.
The large negative $\delta E$ stabilizes a particular deformation.
Hence the energy gaps in the single particle spectrum signal possible stable configurations/deformations.

 In general, energy ${\tilde U}$ depends on the parameters ${\bar \gamma}$ and
 $p$. At the beginning, it seemed that for the method to be meaningful there
 should be certain interval of ${\bar \gamma}$ and corresponding $p$,
 for which ${\tilde U}$ would not practically depend on them (the so called
 "plateau condition"), as it was found for the harmonic oscillator potential.
 However, later it turned out that:

\noindent
 1) The shell correction effectively depends on one parameter:
 ${\bar \gamma}/\sqrt{p}$ \cite{Tajima2010}.

\noindent
 2) For finite-depth potentials the plateau condition usually does not
 hold. In addition, there is a strong and unrealistic dependence of $\delta E$
 on the positive energy spectrum which poses an acute problem for nuclei close
 to the neutron drip line \cite{Nazarewicz1994,Vertse1998}.
 The method can be modified to practically overcome this problem
 \cite{Kruppa1998,Shlomo1997,Vertse2200,Tajima2010}.

\noindent
3) A reasonable condition is that ${\tilde U}$ be close to the semiclassical
 energy, which, after properly taking care for the positive energy spectrum,
 happens for the smallest ${\bar \gamma}>\hbar\Omega$ for which one obtains
 a monotonically increasing smoothed density ${\tilde g}(\epsilon)$
 \cite{Vertse1998,Mohammed2019}.

\noindent
 4) In general, there remains a difference between ${\tilde U}$ and the
 semiclassical energy, typically a few hundred keV, which depends on a nucleus
 and deformation (\cite{Vertse1998,Mohammed2019}).

 For the sake of a complete picture one should add at this point that the
 calculation of the semiclassical density and energy is much more cumbersome
 than that of the Strutinsky shell correction.

 \subsection{Pairing correction}

 The presence of the energy gap in spectra of even-even nuclei and the
 substantially less-than-rigid values of the moments of inertia observed
 in rotational bands are evidence for pair correlations in atomic nuclei.
 Most frequently they are included in theoretical models by adopting
 the BCS formalism from the theory of superconductivity.
 Usually it is assumed that a short-range attraction with a constant
 matrix element $G$ acts between all pairs of time-reversed s.p. states.
The schematic pairing hamiltonian, treated separately for neutrons and for
 protons, may be written as:
\begin{equation}\label{bcs_ham}
H = \sum_{\nu}\varepsilon_\nu a_{\nu}^{+} a_{\nu} -
G \sum_{\nu,\nu^{'}>0}  a_{\nu}^+ a_{\nu^{'}}^{+}
a_{\bar{\nu}^{'}}  a_{\bar{\nu}},
\end{equation}
where, as before, $\varepsilon_{\nu}$ denotes the energy of a single-particle
state $\nu$. Each state $\nu$ has its time-reversal-conjugate
$\bar{\nu}$ with the same energy (Kramers degeneracy).
One assumes the BCS wave function for the nuclear ground state. As it is a
 superposition of components with different numbers of particles, one requires
 that the expectation value of the particle number has a definite value $n$
 ($n=N$ for neutrons and $n=Z$ for protons):
\begin{equation}
\left<\hat{n}\right>=2\sum_{\nu>0}v_{\nu}^2 = n.
\end{equation}
The BCS occupation numbers are given by
\begin{equation}
v_{\nu}^2=\frac{1}{2}\left[ 1-\frac{(\varepsilon_{\nu}-\varepsilon_{ F})}
{\sqrt{\left(\varepsilon_{\nu}-\varepsilon_{ F}\right)^2+\Delta^2}} \right],
\end{equation}
where the parameters $\varepsilon_{ F}$ and $\Delta$ are solutions of the system
of two equations, for the average particle number:
\begin{equation}
\qquad\quad  n = \sum_{\nu>0} \left[1-
\frac{\varepsilon_{\nu}-\varepsilon_{ F}}
{\sqrt{\left(\varepsilon_{\nu}-\varepsilon_{ F}\right)^2+\Delta^2}} \right],
\end{equation}
and for the pairing gap:
\begin{equation}
\label{delta}
\frac{2}{G} = \sum_{\nu>0}
\frac{1}{\sqrt{\left(\varepsilon_{\nu}-\varepsilon_{ F}\right)^2+\Delta^2}}.
\end{equation}
 The larger the number of s.p. states included in the summation
 in Eq. (\ref{delta}) the greater BCS solution for $\Delta$ is
 obtained. This follows from the unphysical assumption of the constant
 matrix element $G$. There are more refined choices of interactions
 producing pair correlations, but the simplest remedy is to fit
 pairing strengths to some data by using a prescribed number of
 included s.p. levels.

 Energy of the system in the BCS state reads:
\begin{equation}\label{bcs_ener}
E_{ BCS} = 2\sum_{\nu>0} \varepsilon_{\nu} v_{\nu}^2
- \frac{\Delta^2}{G} - G\sum_{\nu>0} v_{\nu}^4.
\end{equation}
 For nuclei with odd $Z$ or $N$, one s.p. level is singly occupied.
 This level is excluded - blocked - when the BCS theory is applied to the
 remaining even number of particles.

Pairing correction energy $\delta E^{pair}$
 is constructed in analogy to the shell correction energy $\delta E$,
\begin{equation}\label{bcs_E_pair}
\delta E^{pair} = E_{pair} - {\tilde E}_{pair},
\label{epair}
\end{equation}
where $E_{pair}$ is the pairing energy, i.e. energy difference between the
 paired and unpaired system, corresponding to real
single-particle level distribution $g(\varepsilon)$, and
 ${\tilde E}_{pair}$ is such an energy difference for the smoothed
 single-particle level distribution, ${\tilde g}(\varepsilon)$.
 One has: $E_{pair} = E_{ BCS} - E_{BCS}^{\Delta=0}$,
 where $E_{BCS}^{\Delta=0}$ is the $E_{BCS}$ energy in the
limit of disappearing pairing correlations ($\Delta = 0$),
\begin{equation}
E_{BCS}^{\Delta=0} =
2\sum_{\nu=1}^{n/2} \varepsilon_{\nu} - \frac{Gn}{2}.
\end{equation}
%
The smoothed pairing energy term ${\tilde E}_{pair}$ is usually included in a
 schematic form, resulting from a model with the constant density of doubly
 degenerate energy levels. When calculated in such way,
 it shows nearly no deformation dependence. For example,
it varies by about $50$ keV over the whole deformation range in actinides.
Thus, it could be omitted in energy landscapes, while it shows up in binding
 energies.

The pairing correction (\ref{epair}) counteracts the shell correction $\delta E$;
it is positive where the density of states around the Fermi energy is larger than average
and negative in the opposite case. Therefore a stronger pairing diminishes
the calculated fission barriers.

\subsection{Some remarks on the macro-micro method}

 Concluding this short description of the macro-micro method one can
 notice that the micro and macro parts are to a large extent adjusted
 {\it separately}: the first one to the bulk, and the second
 one to the single-particle properties of nuclei. This makes the method
 theoretically incoherent but more flexible than the HF(B)/functional
 methods in which both types of properties are fixed once the effective
 interaction/functional is chosen. In effect, the fit to nuclear
 properties by selfconsistent models turns out more difficult, for
 example, a good fit to masses does not guarantee realistic fission barriers
 or s.p. levels similar to experimental ones. Moreover, selfconsistent
 calculations are so much more laborious that the systematic
 (that means for many nuclei) and methodologically correct
 (that means: not relying on the energy minimization)
 determination of the fission saddles was for them impossible up to now.

 The Strutinsky method over the last 50 years gave most of the predictions
 concerning nuclear deformations over the periodic table.
  Although it {\it does not} give the exact value of the HF(B) energy,
 the important observation is that, except for very neutron-rich nuclei, there
  were very few, if any, instances in which a macro-micro method based on a
 reasonable phenomenological potential would not reproduce robust
 landscape features predicted by reasonable selfconsistent calculations.

\section{Choice of Models }

  A practical implementation of the macro-micro method will depend on
  choosing a specific model for the macro energy and for the phenomenological
  s.p. potential. Below we describe some most frequent choices.

\subsection{ Macroscopic Energy}

 The expression for nuclear binding energy as a smooth function of $Z$ and
 $N$ is motivated by the saturation property of nuclear forces
 and the idea of {\it leptodermous} expansion. From the experimentally
 measured nuclear masses we see that the binding energy is roughly
 proportional to the number of nucleons $A=Z+N$.
  From the electron scattering off nuclei we learn that
  proton densities of medium-mass and heavy nuclei
 A) show roughly constant surface diffuseness $b\approx 2.4$ fm understood
 as a radial distance between points where density falls from
 90\% to 10\% of its central value and B) have radii following the
 $R\approx r_0 A^{1/3}$ dependence with $r_0\approx 1.2$ fm.
 Hence the nuclear binding coming from the attractive nuclear
 force could be expanded in powers of $b/R\sim A^{-1/3}$ with consecutive
 terms depending on the geometrical properties of nuclear surface $\Sigma$,
 see e.g. \cite{Swiatecki1994}:
 \begin{equation}
 \label{leptoderm}
  E_{nucl}=C_V A - C_s \oint_{\Sigma} dS - C_c\oint_{\Sigma} \kappa dS -
  C_G \oint_{\Sigma} \kappa_G dS + ... ,
 \end{equation}
  where $\kappa$ is the surface curvature, $\kappa=\frac{1}{2}
 (\frac{1}{R_1}+\frac{1}{R_2})$, with $R_1$ and $R_2$ principal radii of
 curvature, and the Gaussian curvature $\kappa_G=\frac{1}{R_1R_2}$.
  For a spherical nucleus these terms are equal to:
 $C_s4\pi R^2\sim A^{2/3}$, $C_c4\pi R\sim A^{1/3}$, and
  $C_G4\pi \sim A^0$. The last term is shape-independent as
 $\oint_{\Sigma}\kappa_G dS$ is the topological invariant.
 The constants in this expansion are in fact
 functions of the neutron excess, $I=(N-Z)/A$, which should be even if we
 accept the charge symmetry of nuclear interactions. One can
 redefine constants so that the new ones will stand by the consecutive powers
 of $A^{1/3}$ multiplied by the corresponding surface integrals divided by
 their spherical values:
  \begin{equation}
 B_S(\Sigma)=\oint_{\Sigma} dS/(4\pi r_0^2 A^{2/3}); \;\;
 B_{curv}(\Sigma)=\oint_{\Sigma}\kappa dS/(4\pi r_0 A^{1/3}).
  \end{equation}

 One also has to add
 the Coulomb energy of proton repulsion, a large term for the heaviest nuclei.
 In the spirit of the expansion (\ref{leptoderm}) its leading term
 is calculated for a constant proton density inside a sharp nuclear surface,
 $\rho_p({\bf r})=\rho_0=-Ze/(4\pi R^3/3)$ inside $\Sigma$, and zero outside
 (with $e$ - the elementary charge):
 \begin{equation}
 E_{Coul}=\frac{1}{2}\int\int \frac{\rho_p({\bf r})\rho_p({\bf r}')
 d^3{\bf r}' d^3{\bf r}} {\mid {\bf r}-{\bf r}'\mid}\rightarrow
 \frac{\rho_0^2}{2}\int\int_{V_{\Sigma}} \frac{d^3{\bf r}' d^3{\bf r}}
  {\mid {\bf r}-{\bf r}'\mid} .
 \end{equation}
  Gathering terms we obtain the liquid-drop formula:
\begin{eqnarray}
\label{mmacr}
 E_{mac}(Z,N,\Sigma)&=&
a_{v}(1-\kappa_{v} I^2)A -a_{s}(1-\kappa_{s}
I^2)A^{2/3}B_S(\Sigma) \nonumber\\
&&- a_c(I^2)A^{1/3}B_{curv}(\Sigma) - a_0 A^0 \\
&& - c_1 Z^2 A^{-1/3} B_C(\Sigma) + c_4 Z^{4/3}A^{-1/3} + ... , \nonumber
\end{eqnarray}
 where $c_1=\frac{3}{5} \frac{e^2}{r_0}$, and
 $B_C(\Sigma)$ - a shape-dependent ratio of dimensionless Coulomb energies
 for the deformed and spherical nucleus:
  \begin{equation}
 B_C(\Sigma)=
 \frac{15}{32\pi^2 R^5}\frac{\int\int_{V_{\Sigma}} d^3{\bf r}'d^3{\bf r}} {\mid {\bf r}-{\bf r}'\mid}.
  \end{equation}

  The last term in expression (\ref{mmacr}) is the
 quantal Coulomb-exchange term for a spherical nucleus in the Slater
 approximation. It is an example of deformation-independent terms that may be
 also included in the liquid-drop energy in order
 to improve the agreement with the experimental binding energies by
  accounting for some physical effects.
 The volume energy term is the largest, but since it is constant for a given
 nucleus the crucial shape dependence of $E_{mac}$ resides
  in the functions $B_{S}$ and $B_{C}$, describing the
  surface and Coulomb energies, respectively.
  They are crucial for the fission barriers and determine them
  nearly exclusively in medium-heavy nuclei, in which the macroscopic
 shape-dependent terms dominate the shell correction.
 The surface energy grows with deformation while the Coulomb energy decreases.
  The deformation-dependent curvature term turned out close to zero in some
 adjustments of models to experimental masses, but it is present in the
 LSD macroscopic formula of \cite{Dudek2003}.

 There is another approach to macroscopic energy in which the surface and
 Coulomb energies from the outset include corrections for the surface
 diffuseness and/or finite range of interaction. A general way is to present
 them as double folding:
 \begin{equation}
 \label{folding}
   E = \int\int d^3{\bf r}d^3{\bf r}'  \rho({\bf r})\rho({\bf r}')
   f({\bf r}-{\bf r}') ,
 \end{equation}
 where $\rho$ are nucleon (or proton, for the Coulomb energy) densities
 and the function $f$ models the interaction with a finite range. One
 can induce diffuseness in a sharp density, equal $\rho_0$ within the volume
 $V_{\Sigma}$, by folding it with a localized profile, like, for example,
  the Yukawa function:
 \begin{equation}
  g_Y(r_{12})=\frac{1}{4\pi a^3}\frac{e^{-r_{12}/a}}{r_{12}/a}  ,
 \end{equation}
 so that $\rho({\bf r}_1)=\rho_0\int_{V_{\Sigma}} d^3{\bf r}_2\
 g_Y(|{\bf r}_1-{\bf r}_2|)$,
 where $r_{12}$=$|{\bf r}_1-{\bf r}_2|$, $a$ is the width of the
 Yukawa profile, and $\int d^3{\bf r}\ g_Y({\bf r})=1$. Using $g_Y$ as the
 nuclear interaction $f$ and sharp surface densities one can model surface
 energy after subtracting form (\ref{folding}) the volume term. This approach
 was modified in \cite{Krappe1979} to the "Yukawa plus exponential" model which does not require any subtraction.
  The function $B_S(\Sigma)$ in this model is defined as:
\begin{equation}
\label{B_S}
B_S(\Sigma) = \frac{A^{-2/3}}{8\pi^2r_0^2a^4}\int\int_{V_{\Sigma}} \left(
2-\frac{r_{12}}{a} \right) \frac{e^{-r_{12}/a}} {r_{12}/a} d^3r_1
d^3r_2
\end{equation}
  and was used in the Finite Range Liquid Drop Model (FRLDM)
  \cite{Moller1981,Moller1995,Moller2016}.
 For charge densities folded with $g_Y$
 one can obtain from Eq. (\ref{folding}) the expression
 for the Coulomb energy \cite{Davies1977}:
\begin{equation}
\label{B_C}
B_C(\Sigma) = \frac{15}{32\pi^2} \frac{A^{-5/3}}{r_0^5}\int\int_{V_{\Sigma}}
\frac{1}{r_{12}} \left[ 1- \left(
1+\frac{1}{2}\frac{r_{12}}{a_{ den}} \right) e^{-r_{12}/a_{den}} \right]
 d^3r_1 d^3r_2 ,
\end{equation}
also used in FRLDM.
In the above formulas, $r_{12}$=$|{\bf r}_1-{\bf r}_2|$, with
${\bf r}_1$ and ${\bf r}_2$ the
positions of interacting volume elements, $a$ - the range of
the Yukawa interaction, $a_{den}$ -
 the range of the Yukawa function used to generate nuclear
charge distribution. The functions are normalized in such a way
that they are equal 1 for a spherical nucleus in the limit
 of $a$=0 (for $B_S$) and $a_{den}$=0 (for $B_C$),
 corresponding to the traditional liquid-drop model with a sharp
 surface.

\subsection{ Phenomenological Deformed Potential}

 The phenomenological s.p. potential used to calculate shell correction
 energy is usually adjusted to the s.p. levels experimentally observed
 around Fermi energy in various nuclei. In order to reproduce observed
 magic shells it contains the spin-orbit part in addition to the central
 one and there is Coulomb potential for protons.
 Hence its general form may be written as follows:
\begin{equation}\label{micr_pot}
V_{micr}=V_{centr}+\lambda\left(\frac{\hbar}{2mc}\right)^2
\left( \nabla  V_{s.o.}\right) \cdot
\left(\vec{\sigma}\times\vec{p}/\hbar\right)+V_{C},
\end{equation}
 with the central potential $V_{centr}$, the spin-orbit potential
 $V_{s.o.}$, with $m$ - nucleon mass, the s.o. strength $\lambda$,
  and the Coulomb potential for protons $V_C$. In general, the central
 and s.o. potentials for neutrons and protons are different.

 Deformation of a nucleus is usually introduced to the s.p. potential
 in a similar way as in the macroscopic part - through a definition of
 the nuclear surface $\Sigma$ via the equation involving deformation
 parameters, collectively designed $\beta$.
 The mostly used in the macro-micro methods are the
 folded Yukawa and the Woods-Saxon potentials.

 The central part of the folded Yukawa potential is generated by the function
 $g_Y$ with some chosen range $a_p$:
 \begin{equation}
 V_Y({\bf r})=-\frac{V_0}{4\pi a_p^3}\int_{V_{\Sigma}}
 \frac{e^{-|{\bf r}-{\bf r}'|/a_p}}{|{\bf r}-{\bf r}'|/a_p} d^3{\bf r}' .
 \end{equation}
 The spin-orbit potential is the same as the central one, $V_{s.o.}=V_Y$.
 The potential depth $V_0$, its radius defining
  the volume enclosed by the surface $\Sigma$ and the spin-orbit strength
 $\lambda$
 are adjusted to data and depend on $Z$ and $N$
 \cite{Moller1995}. The Coulomb potential for protons is
 calculated for a uniform, sharp density $\rho_0=Ze/(4\pi A r_0^3/3)$.

 The Woods-Saxon potential which we use in our macroscopic - microscopic
 model is defined as follows. Its central part as given
 by S. \'Cwiok et al \cite{WS} reads:
\begin{equation}\label{ws_pot}
V_{WS}(\vec{r})=-\frac{V}{1+e^{d(\vec{r},{\beta})/a_{ws}}},
\end{equation}
where $V$ is the potential depth, $d(\vec{r},{\beta})$ is
the distance from the point $\vec{r}$ to the surface of the
nucleus $\Sigma$, taken with the plus sign outside, and minus inside
 $\Sigma$, $a_{ws}$ is the diffuseness of the nuclear surface.
 For a spherical shape ($\beta=0$), the function
 $d(\vec{r},{\beta})$ reduces to: $r-R_0$ with a constant radius of a nucleus
 $R_0 = r_0 A^{1/3}$. The depth of the potential is
\begin{equation}\label{ws_pot_depth}
V = V_0 (1 \pm \kappa I),
\end{equation}
where $I=(N-Z)/A$, and $V_0$ and $\kappa$ are adjustable parameters.
 The sign ($+$) is for protons and ($-$) for neutrons.

The full W-S potential has the form (\ref{micr_pot})
%
%
 with the spin-orbit part of the same form as $V_{WS}$, but with a
 different radius parameter. The radii and the s.o. strenghts are the
 parameters of the model. The Coulomb potential is:
 \begin{equation}\label{coul_pot}
V_{C}(\vec{r})=\rho_{c}\int_{V_{\Sigma}}\frac{d^3r'}{|\vec{r}-\vec{r'}|},
\end{equation}
where $\rho_{c}=(Z-1)e/(4\pi R_0^3/3)$ is the uniform charge density and the
integration extends over the volume enclosed by the nuclear surface.

\section{ Search for Saddles }

  There are some points relevant to the application of the macro-micro
  method which are independent of the choice of the particular macro or
 micro model. Among them are those related to a search for fission saddles.

 i) Which deformation parameters to use?

 An arbitrary shape of a drop requires infinite number of coordinates.
 However, those of small wavelengths, close to the size of a nucleon, should
   be discarded as unphysical. Unfortunately, a reasonable cut off will depend
  on particular set of employed coordinates.
  Rather complete review of various deformation parametrizations
  used in studies of fission is given in \cite{Hasse1988}.
  The most used ones are: the parametrization by 3 matched quadratic
  surfaces and the one used here - the coefficients of expansion onto
  spherical harmonics. The first one is especially useful for elongated shapes
  close to scission and used by P.~M\"oller et al to describe
  axially-symmetric shapes. The second one, used in our model,
  is general, but we avoid shapes without a symmetry plane.
 The nuclear surface $\Sigma$ is defined
 by the equation for the surface radius as a function of spherical angles
 $\vartheta$ and $\varphi$ \cite{WS}:
  \begin{eqnarray}
  \label{shape}
    R(\vartheta,\varphi)&=& c(\{\beta\}) R_0 \{ 1+\sum_{\lambda\ge 1}
 \beta_{\lambda 0} Y_{\lambda 0}(\vartheta,\varphi)+
\nonumber \\
&&
    \sum_{\lambda\ge 1, \mu>0}
   \beta_{\lambda \mu } Y^r_{\lambda \mu} (\vartheta,\varphi)\}  ,
\end{eqnarray}
  where $c(\{\beta\})$ is the volume-fixing factor. The real-valued spherical
  harmonics $Y^r_{\lambda \mu}$ are defined in terms of the usual ones as:
 $Y^r_{\lambda \mu}=(Y_{\lambda \mu}+Y_{\lambda -\mu})/\sqrt{2}$ for $\mu$
 even, and:
 $Y^r_{\lambda \mu}=-i(Y_{\lambda \mu}-Y_{\lambda -\mu})/\sqrt{2}$ for $\mu$
  odd. In other words, we can treat shapes with one symmetry plane $yz$.
 Usually, however, we deal with shapes having two symmetry planes for which
 $\beta_{\lambda \mu}=0$ for $\mu$ odd.

 ii) How many deformations are sufficient?

  At present, the practical restrictions by available computation resources
 confine one to 5 - 6 coordinates when performing calculations for many
 nuclei, which is about the minimal number
   necessary to describe shapes from the ground state to beyond the
  second barrier.
  This practically forces one to use different parametrization when
 describing ground states and different saddles.

 iii) How to represent many-dimensional landscapes?

\begin{figure}[h]
\vspace{-0mm}
\hspace{4mm}
\centerline{\includegraphics[scale=1.00]{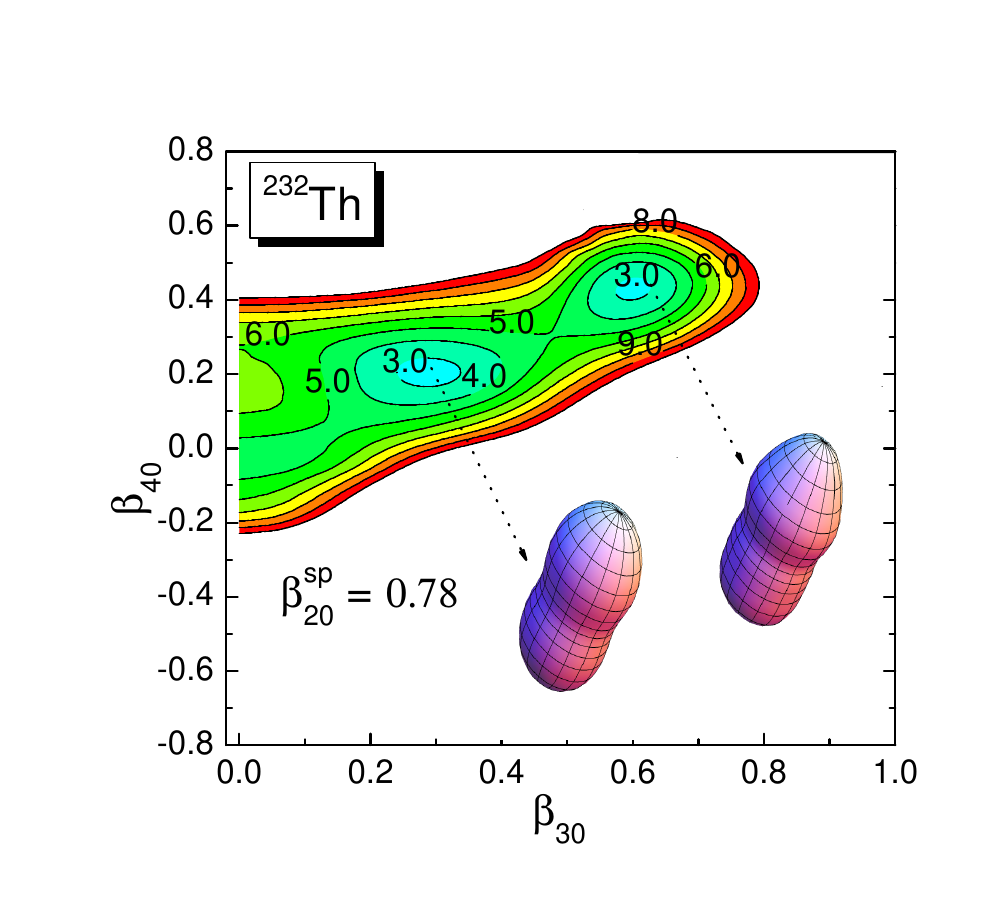}}
\vspace{-0mm}
\caption{{\protect Fragment of the energy surface for $^{232}$Th
 projected on the ($\beta_{30}$,$\beta_{40}$) plane by fixing
 $\beta_{20}=0.78$ and minimizing over
 $\beta_{50}$,$\beta_{60}$,$\beta_{70}$,$\beta_{80}$; the minima correspond
 to shapes lying slightly beyond the second saddle.
 Taken from \cite{Jachimowicz2012}; copyright 2012 by APS. }}
\label{b3b4Th232}
\end{figure}

 Any $n$-dimensional landscape with $n\geq 3$ is difficult to visualize.
 For a long time, two-dimensional energy maps serving the search for fission
 saddles were obtained by minimizing over the remaining, i.e. over the other
 $n-2$, shape coordinates. This was an obvious and easiest choice, especially
 in the selfconsistent studies, which are based on the minimization with
 constraints. Unfortunately, such maps can distort reality if there are
  multiple minima in those $n-2$ coordinates. An example is provided in
  Fig. \ref{b3b4Th232}. It follows that so obtained
  maps may be misleading as to the situation od fission saddles.

 iv) How to search for saddles?

  Saddles on the energy landscape are neither minima nor maxima, therefore
  a proper method of search for them is required, different from
  the minimization.
  The fallacy of the minimization method with constraints in finding
 fission saddles was discussed on examples in \cite{Myers1996,Moller2000} and
 \cite{Moller2009}.
 The related errors are also present
 in selfconsistent HFB or HF+BCS studies, see e.g. \cite{Dubray2012},
  where their elimination would require much more laborious calculations.

   Two of possible correct methods are the "Dynamic Programming Method"
 used by A. Baran in the context of finding the subbarrier least action
 trajectory \cite{Baran1978,Baran1981} and the "Imaginary Water
  Flow" (IWF) borrowed from geographical topography studies, first used
  for finding fission barriers in \cite{Mamdouh1998}. Both start from
  calculating energies on a n-dimensional grid of deformations, of which one,
 say the first, $q_1$, controls the elongation of the nucleus.
 Points $(q_1,q_2,...,q_n)$,
 with $q_1$ fixed and other $q_i$ taking all possible values,
 form hypersurfaces enumerated by increasing values of $q_1$.
 The first method aims at minimizing some functional over paths in consecutive
 steps:
 keeping the calculated optimal functional values for each point of the
 hypersurface $k$ it updates the optimal values for each point of the
 next hypersurface $k+1$. If one chooses the maximal energy on a path
  $V_{max}({\rm path})$ as the functional, then its minimization over paths
 will give the fission barrier, see (\ref{barr1}).
  In the second method, one simulates a gradual filling of the minimum with
 water and looks for a point at which the water starts to overflow to the
 neighbouring valley.
 Both methods require rather small mesh size and large memory,
 but the second one is more efficient and can treat up to six-dimensional
 grids in systematic calculations (8D for a single nucleus).
 Both have to be properly modified if there are many saddles - each of
 the saddles should be found and a selection procedure must be applied to
 choose the one defining the fission barrier height.

 \section{Specifications of the Model}


%
%

 The results which will illustrate some applications of the macro-micro method
 were obtained, unless stated otherwise, within the model specified in the
 recent work \cite{Jach2021}. The macroscopic energy
 included surface and Coulomb terms given by Eq. (\ref{B_S},\ref{B_C}),
 and parameters specified in \cite{Jach2021}.
 The volume integrals in $B_S$ and $B_C$, after turning them into surface
 integrals, were calculated
 by using the four-fold (or three-fold, for the axially symmetric shapes)
 $N$-point Gauss quadrature with $N\ge 64$.
  The deformed W-S potential was used as the phenomenological mean field,
 with the ''universal'' set of parameters given in \cite{WS}.
 The s.p. energies were calculated by using $n_{p}=450$ lowest proton and
 $n_{n}=550$ neutron levels from $N_{max}=19$ lowest shells of the deformed
 harmonic oscillator.
 The shell correction was calculated with the smoothing parameter
 ${\bar \gamma}=1.2\times 41/A^{1/3}$ MeV
  and the sixth-order polynomial ($p=6$) for $f_p$.
  The pairing strengths: $G_n=(17.67-13.11\cdot I)/A$ for neutrons,
 $G_p=(13.40+44.89\cdot I)/A$ for protons ($I=(N-Z)/A$),
  as adjusted in \cite{WSpar}, were used when solving the BCS equations
  including the lowest $N$ (doubly degenerate) neutron and
  $Z$ proton levels.
  The same parameters of the model were used previously in all our
 calculations of masses \cite{Jach2021} and fission barriers
 \cite{Jachimowicz2017_I,Jachimowicz2017_III} of actinides and
 superheavy nuclei.

  For odd and odd-odd nuclei, at each deformation, the energy minimization
 was performed over configurations of the odd particle,
 from the 10-th s.p. level below to the 10-th above the Fermi level.
 Pairing was included using the BCS with blocking.
 This means that we were looking for adiabatic saddle points in these nuclei.

  After energy was calculated on a 5 - 7 dimensional deformation grid
 for each nucleus (see next section), the IWF procedure was applied to
 find fission saddles and select those which determine barrier heights.


\section{Fission Barriers in Actinides}

 Fission barriers in actinide nuclei, which result from addition of
 deformation-dependent macro and micro terms of similar magnitude,
   exhibit a two-hump structure
 with both humps $~$5-6 MeV high in U or Pu, with the second one decreasing
 with $Z$, $B_B({\rm Th})-B_B({\rm Cf})>2$MeV. Their structure was revealed
 by the discovery of fast fissioning
 states by Polikhanov et al. \cite{Polikhanov1962} and explained within the shell correction
 method by Strutinsky \cite{Strutinsky1969}. The explanation is that these are the ground,
 and sometimes also excited, states in the second well at large deformation
 and energy $E^*_{II}$ above the g.s., and therefore protected from fission
 only by the second barrier reduced by $E^*_{II}$, $B_B-E^*_{II}$. Hence
 the name: shape- or fission isomers. The short
 fission half-lives of 10 ps - 10 ms make their study quite difficult.

 The decrease in the second barriers has the exponential effect on
 spontaneous fission rates/half-lives, for example:
 $T_{1/2}^{sf}\approx 10^{19}$ yr for $^{235}$U, $10^{14}$ yr for $^{241}$Am,
 86 yr for $^{252}$Cf and 8 s for $^{252}$No. At the same time, the height of
 the fission barrier in these nuclei is similar to the energy of separation of
 one neutron. Some actinide isotopes are fissile - they fission after
 a capture of thermal (i.e. with nearly zero kinetic energy) neutron.
 Applications of this phenomenon, first military but later also some peaceful,
 triggered many experimental studies of the induced fission which provided the
 current knowledge about the parameters of the barriers in actinides.

 It is important to mention that the method of extracting heights and other
 parameters of the fission barriers from experiments is based on the
 one-dimensional picture of Fig. \ref{schem}, see e.g. \cite{Capote2009}.
  This adds to the uncertainty of comparisons between the quantity $B$ defined above in terms
 of the multi-dimensional model and the "experimental" values reported usually
 as a result of fitting the excitation function in the neutron-induced
 reaction. One has also to mention that two sets of evaluated fission barriers
 in actinides differ for some nuclei by more than 0.5 MeV.


\begin{figure}[h]
\vspace{-0mm}
\hspace{0mm}
\centerline{\includegraphics[scale=0.4,angle =-90]{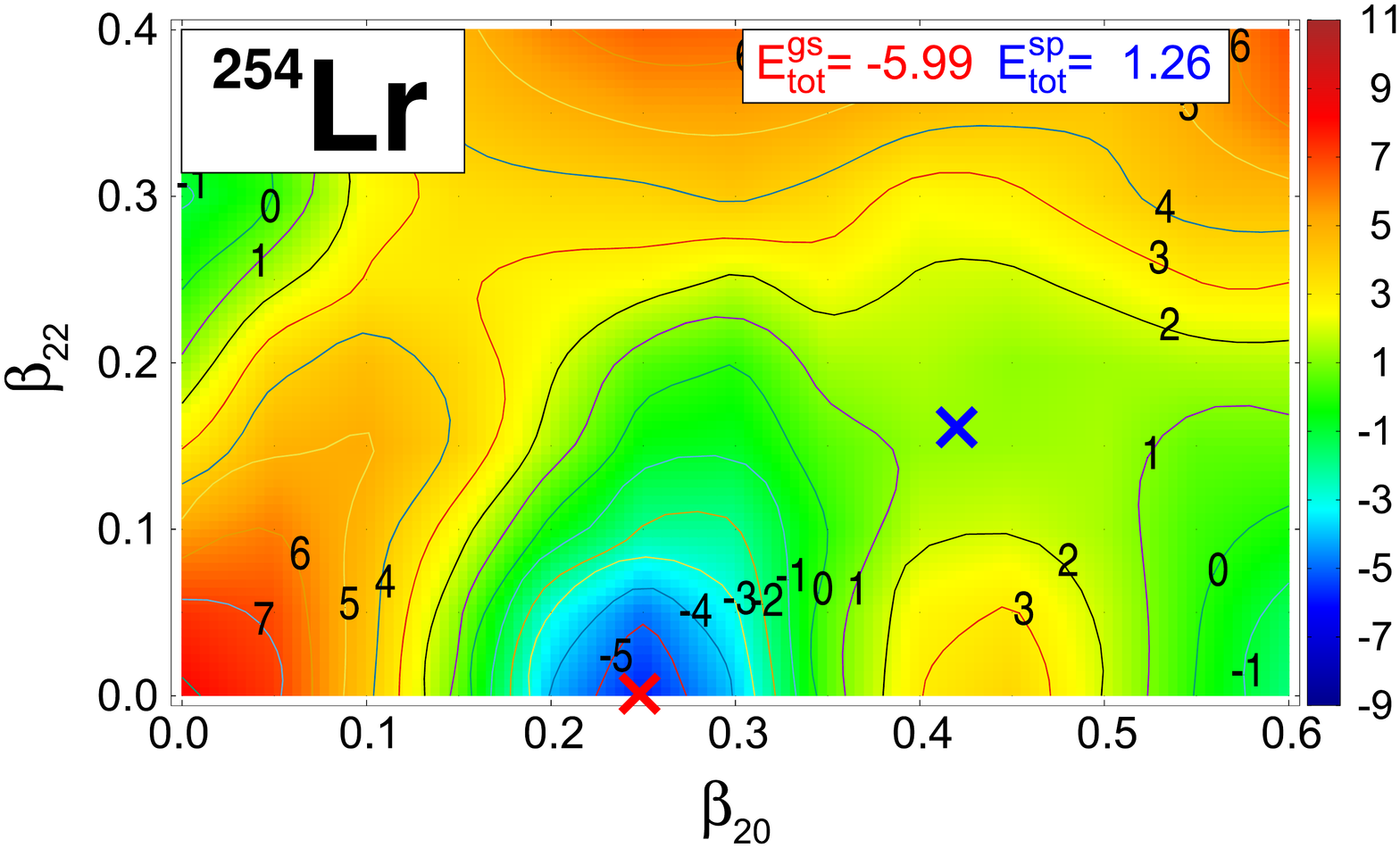}}
\vspace{-0mm}
\caption{{\protect Energy map in ($\beta_{20}$,$\beta_{22}$) plane for
  $^{254}$Lr obtained by minimization over three remaining
 shape parameters $(\beta_{40},\beta_{60},\beta_{80})$;
 energy (in MeV) is relative to the macroscopic energy at the
 spherical shape.}}
\label{254Lr}
\end{figure}

\begin{figure}[h]
\vspace{-0mm}
\hspace{0mm}
\centerline{\includegraphics[scale=0.6]{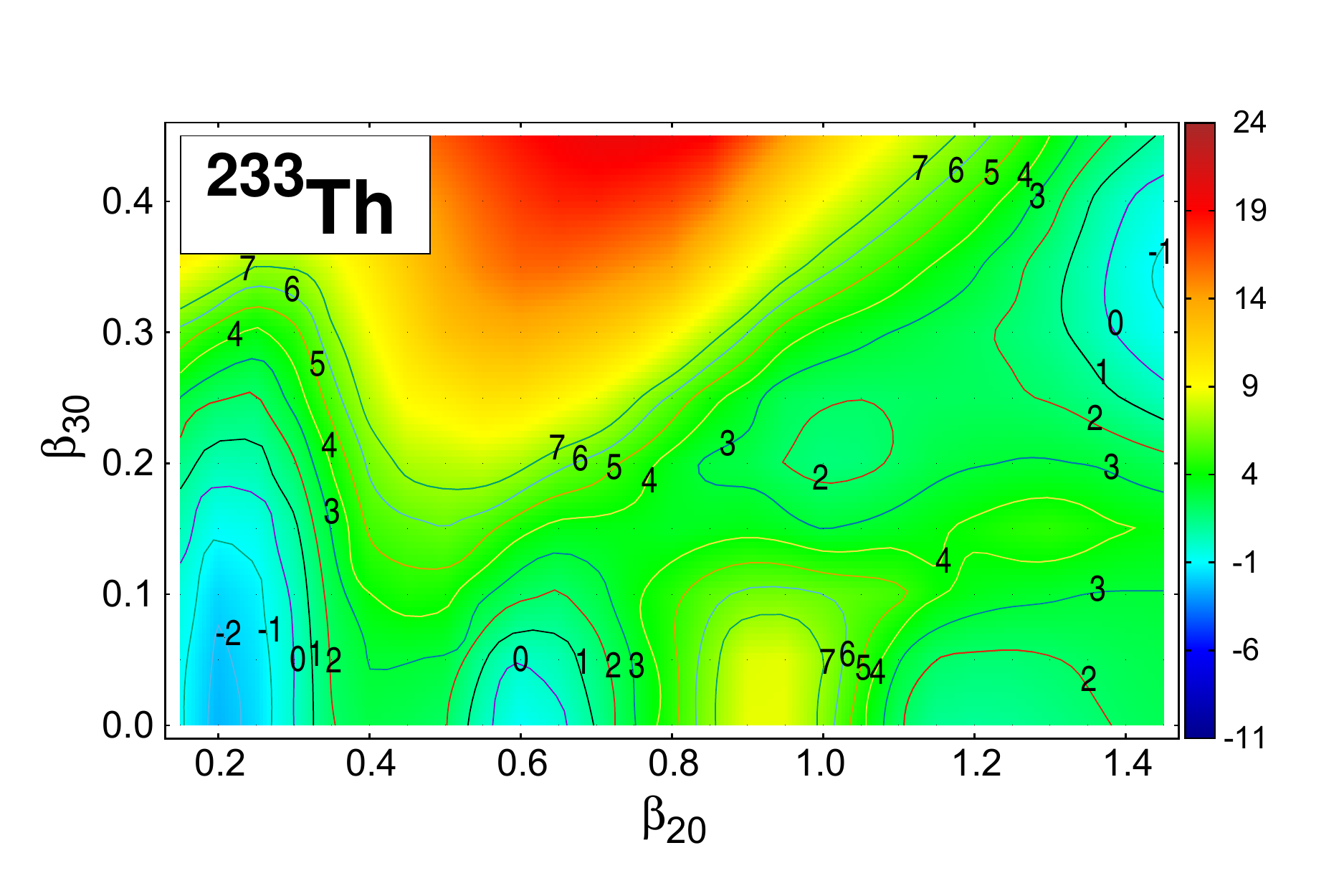}}
\centerline{\includegraphics[scale=0.6]{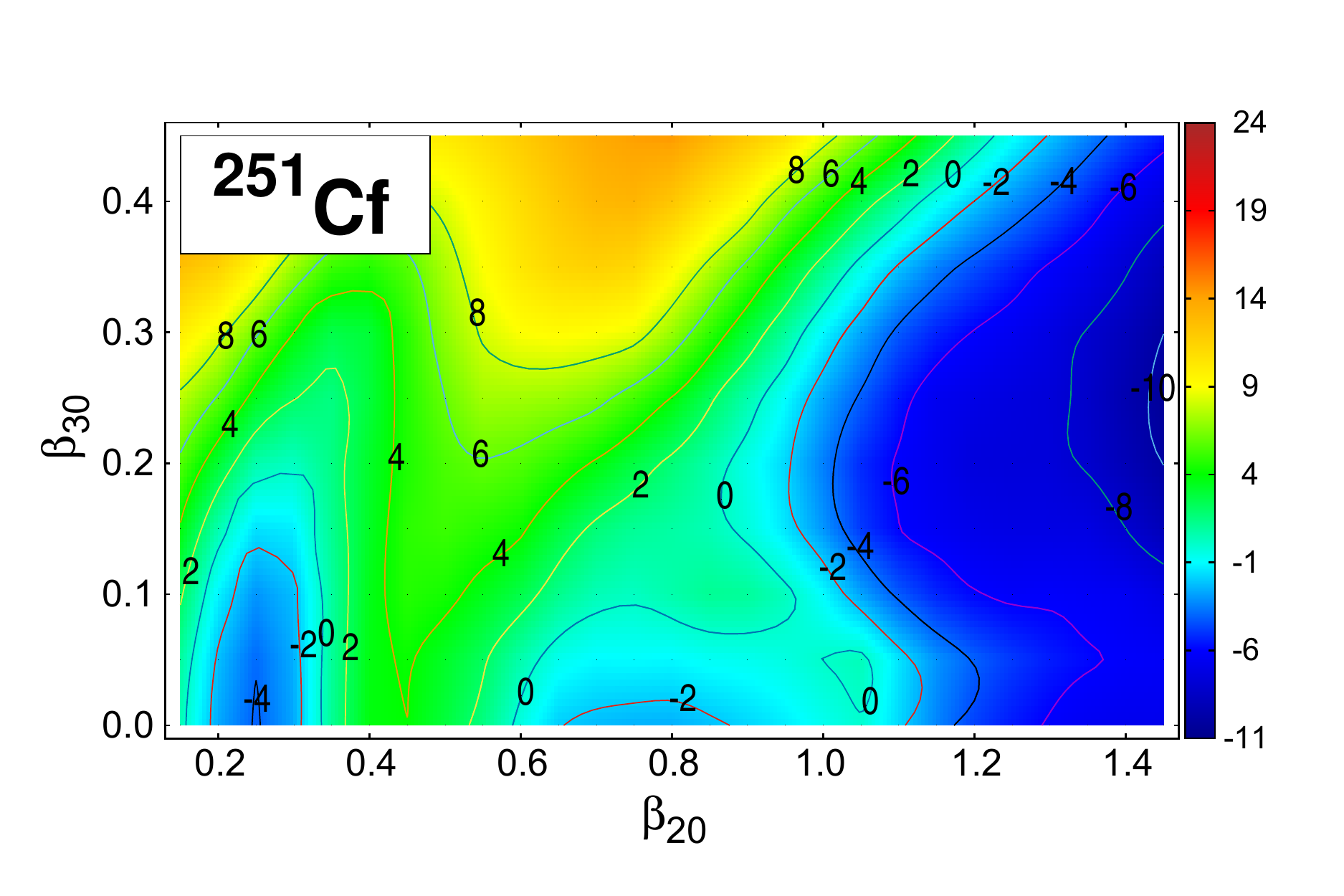}}
\vspace{-0mm}
\caption{{\protect Energy maps in ($\beta_{20}$,$\beta_{30}$) plane for
 $^{233}$Th and $^{251}$Cf obtained by minimization over five remaining
 shape parameters $\{\beta_{40},\beta_{50},\beta_{60},\beta_{70},
 \beta_{80}\}$ with $\beta_{10}$ fixed at each point by the center-of-mass
 condition; energy (in MeV) is given relative to the macroscopic energy at the
 spherical shape. Taken from \cite{Jachimowicz2017_III}; copyright 2020 by APS.
  }}
\label{ThCf}
\end{figure}

 As followed from numerous studies \cite{Thirolf, Krasznahorkay2011}, shapes of actinide nuclei
 characteristic of primary (g.s.) and secondary (shape-isomeric) minima
 and fission saddles belong to different symmetry classes.
  The minima are axially symmetric, with reflection-asymmetric shapes showing
 only in ground states of very light isotopes.
 They  were found by performing the energy
  minimization over deformations $\beta_{20}$-$\beta_{80}$ using the
  conjugate gradient method.
  To avoid falling into local minima, the minimization was repeated dozens
  of times for each nucleus with randomly selected starting deformations.
  For odd-$Z$ or/and odd-$N$ systems, the minimization over configurations
  was performed at every step of the gradient procedure.

  The comparison of calculated and measured excitation energies $E^*_{II}$
  of isomeric minima in 28 actinide nuclei, mostly Pu, Am and Cm isotopes,
  shows that our calculation underestimates them by 0.3 MeV on average,
 with the root-mean-square (rms) deviation of 0.53 MeV, and the largest
 difference of 1.1 MeV. One has to notice that the measured values have often
 uncertainty of 0.2 - 0.4 MeV. The deformations $\beta_{20}$ of secondary
  minima are between 0.5 and 0.80 and grow with $A$.


\subsection{First and Second Barriers}

 One has to note that deformations of the first and even second fission
 saddles in actinides, although much larger than in their ground states,
 are still far
 from that of the scission configuration of two nearly disjoint fragments.
 Therefore, one can hope to describe them in terms of a relatively
 limited number of deformations $\beta_{\lambda\mu}$.
 When reaching behind the second barriers, one should check the reliability
 of the applied shape variety for the determination of the energy
 landscape, as the search for the third minima illustrates (see below).

 The first barriers are very often considerably lowered by the nonaxiality \cite{Pashkevich1969, Larson1972,Schultheiss1971,Larson74}.
 This is illustrated in Fig. \ref{254Lr}, where we show the energy map
 in $(\beta_{20},\beta_{22})$ plane for $^{254}$Lr. One can see a substantially
 triaxial first saddle marked by the blue cross.
 On the other hand, the second barriers are considerably lowered by
 the reflection asymmetry - the fact which allowed Strutinsky to explain the
  puzzle of asymmetric fission of actinides \cite{Moller70,Moller72}. Two examples are shown
 in Fig. \ref{ThCf}, where energy maps in $(\beta_{20},\beta_{30})$ plane
 are shown for $^{233}$Th nad $^{251}$Cf. One can see mass - asymmetric
 second saddles, a more pronounced in Th.
 This knowledge suggests using different deformation sets while searching
 for the first and the second barriers.

  The first saddle points were searched by the IWF
 algorithm on the five-dimensional grid including even-$\lambda$ deformations
 $\beta_{20}$-$\beta_{80}$, within the range of $\beta_{20}$: 0 - 0.60, and
 the quadrupole nonaxiality $\beta_{22}$ - altogether 5 variables. The initial
 grid of 29250 points was
 fivefold interpolated to more than 50 million points to perform the IWF.
 The grid for the search of the second saddles used the same deformations as
  for minima, but for each point $\beta_{20}$-$\beta_{80}$ (seven variables)
 the deformation $\beta_{10}$ was added,
 determined by the center - of - mass condition. With $\beta_{20}$ reaching
 1.5, the seven-dimensional mesh contained $\approx7.5$ million points.
 As a two-fold interpolation did not modify the results of the IWF search
 we decided to find second saddles without any interpolation in this case.

\begin{figure}[h]
\vspace{0mm}
\hspace{0mm}
\includegraphics[scale=0.45]{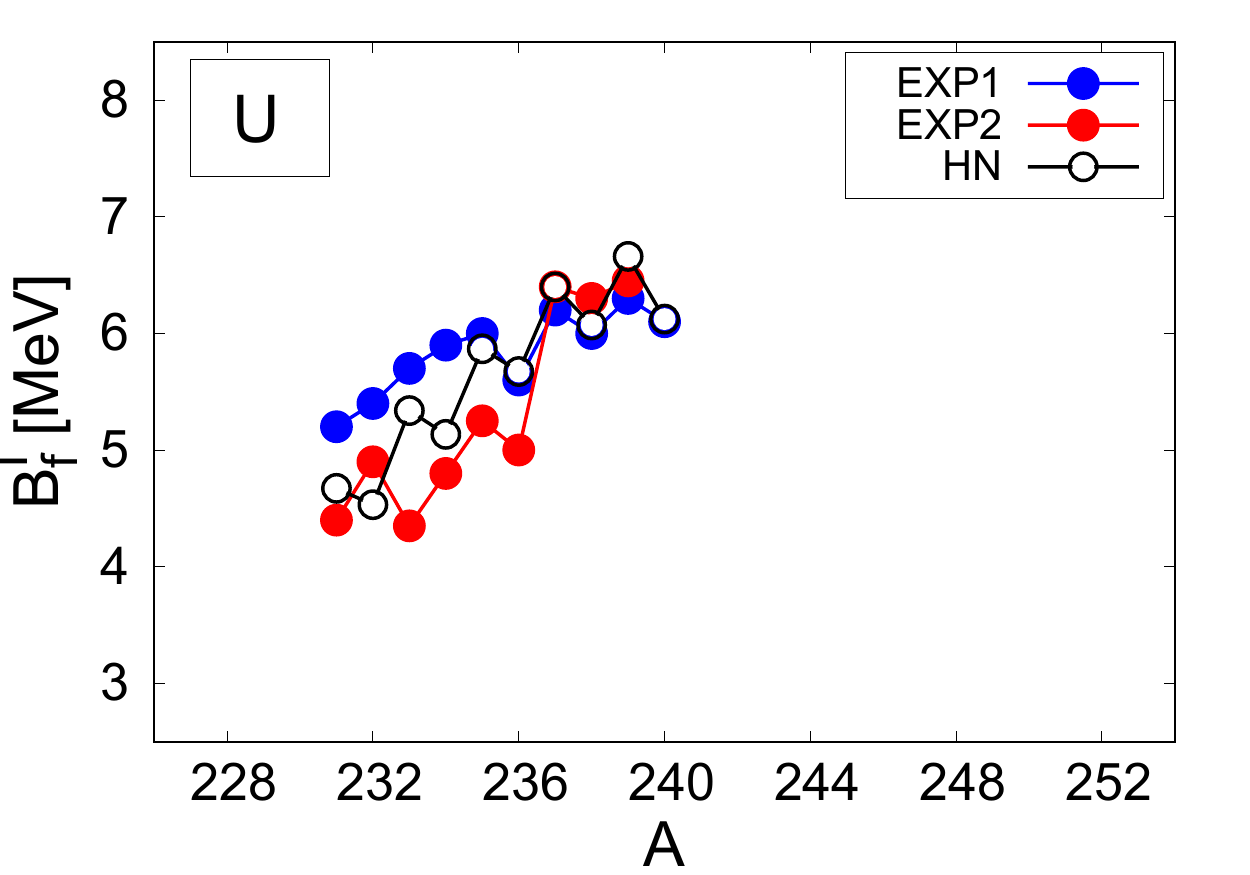}
\includegraphics[scale=0.45]{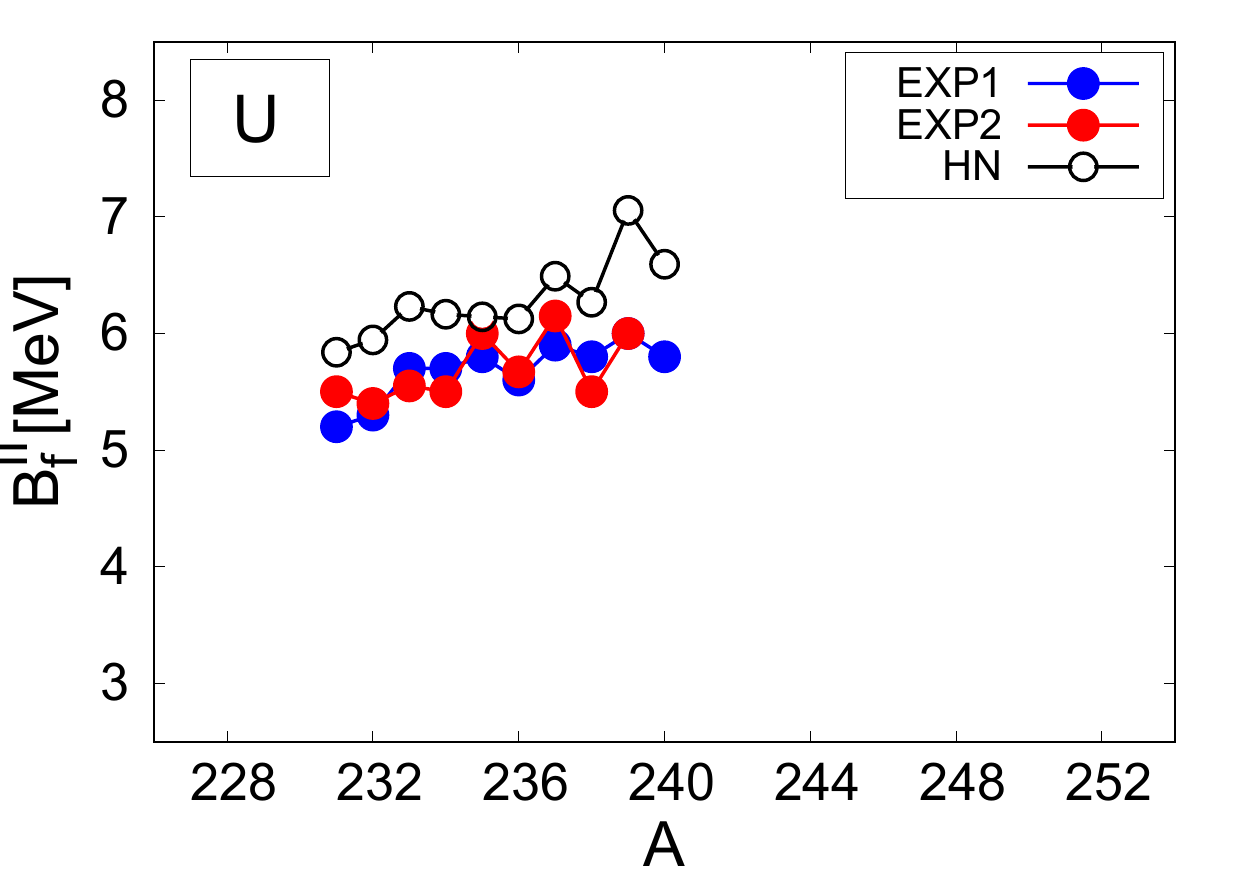}
\includegraphics[scale=0.45]{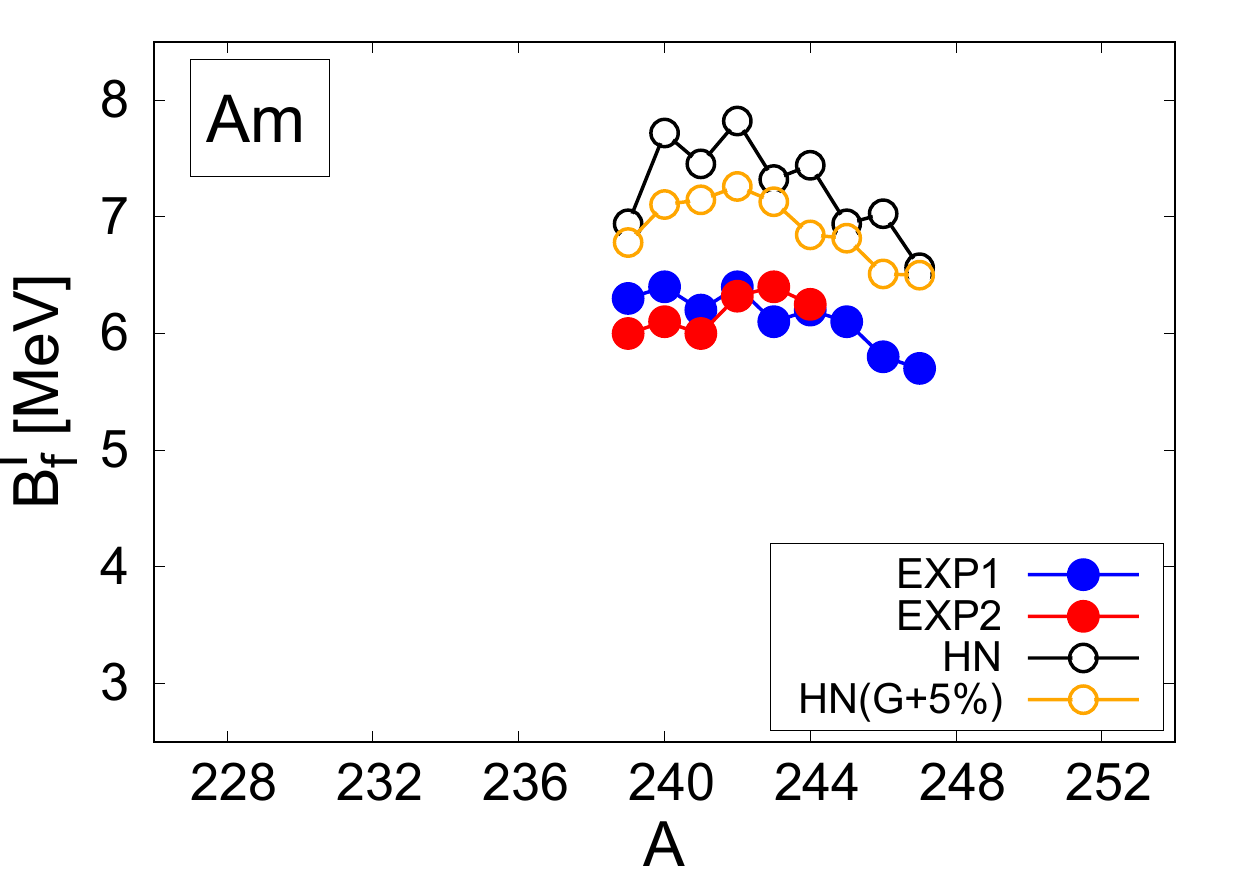}
\includegraphics[scale=0.45]{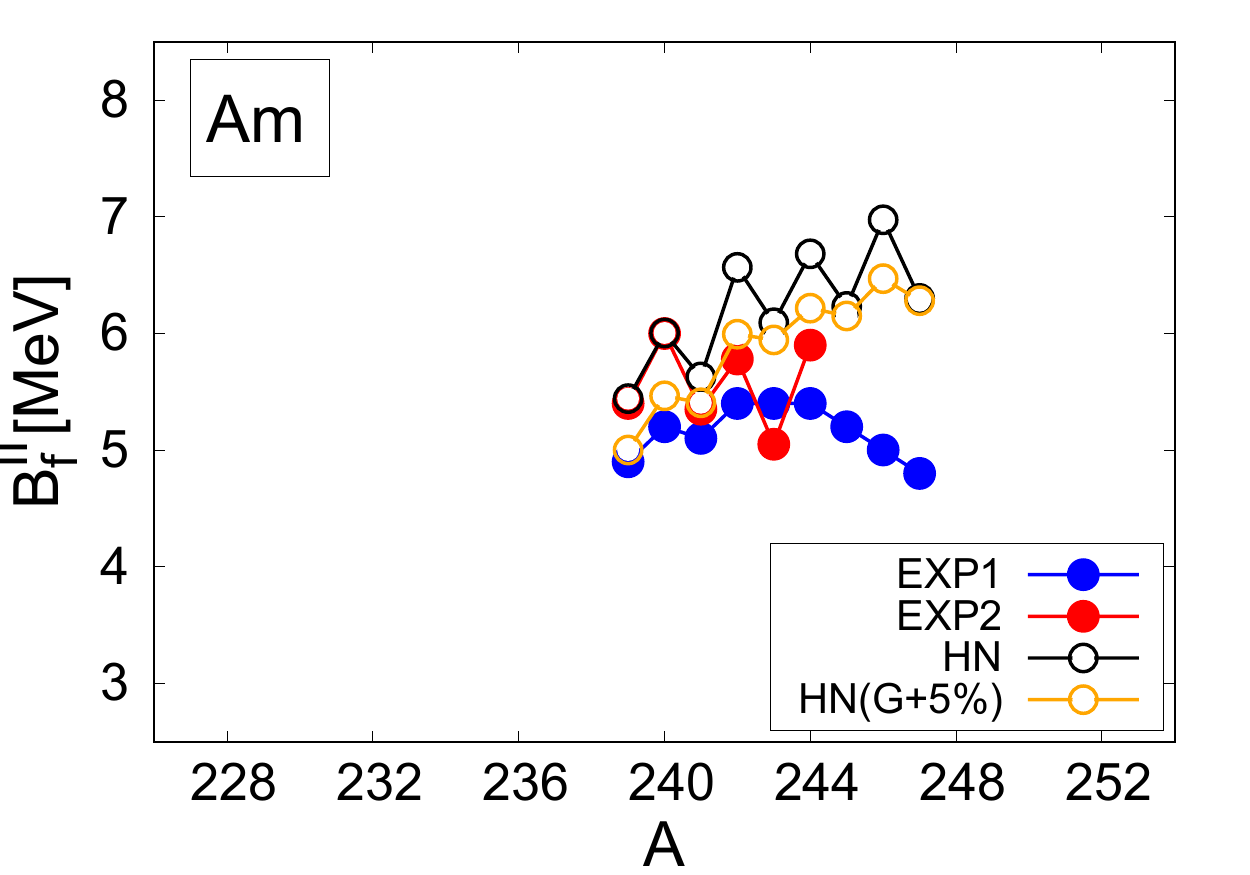}
\vspace{-0mm}
\caption{{\protect First and second fission barrier heights in uranium and americium
 nuclei: results of the present model (black circles) and experimental
 (G. N. Smirenkin 1993) (red dots) and (Capote et al 2009) (blue dots).
 For americium nuclei, the effect of $5 \% $ enhanced pairing was additionally shown (open orange circles)
 Taken from \cite{Jachimowicz2017_III}; copyright 2020 by APS. }}
\label{BfI-BfII}
\end{figure}


  The calculated first fission barriers are substantially too small for Th
 isotopes, close to the values evaluated from experiments for Pa and U, and
 too large on average for larger $Z$. The first effect, known as the
 "thorium anomaly" is common to many theoretical calculations and, perhaps,
  related to the barrier shapes different from those assumed in the analysis
 of experimental data.
  The second barriers, except for Th and the lightest isotopes of
  Pu and Cm, come out overestimated. Additionally, those for Pu and Am
  show an increase with $N$ which does not exist in experiment, but
  shows up in many selfconsistent calculations.
  Both, the first and second barriers in our model show the
  odd-even staggering which is more pronounced than in experiment, to a degree
  which depends on which experimental set one uses for comparison.
   In Fig. \ref{BfI-BfII} we present the calculated barriers vs experimental
 data for "well" and "not-so-well" agreeing with the data isotopic U and Am
 chains. The agreement with the "experimental" barriers for Am isotopes may
 be slightly
 improved by assuming larger by 5\% pairing strength for odd-$Z$ and odd-$N$
 nuclei. The corresponding barriers, calculated for the sake of test,
 are shown in orange in Fig. \ref{BfI-BfII}. Such change would slightly spoil
 the agreement of calculated nuclear masses with the experimental ones.

 The present ability of the MM model to reproduce barrier heights in
 actinides may be illustrated by the obtained root-mean-square deviations from
 two existing sets of experimental estimates, including respectively 71 and
 45 nuclei. For the first barriers they are: 0.94 (set I) and
 0.85 MeV (set II),
 for the second barriers: 0.92 (set I) and 0.82 MeV (set II).
 Within the FRLDM of M\"oller et al \cite{Moller2009} these deviations are
 quite similar for the second barriers, but sizably larger for the first
 [1.48 (set I) and 1.35 MeV (set II)].

\subsection{Uncertain Third Minima and Barriers}

 Fission probabilities in neutron capture reactions on some actinides
 show narrow resonances at excitation energies above the second barrier.
 A fine structure of some was interpreted as a signature of multiple
  hyperdeformed (i.e. having moments of inertia larger than in the isomeric
 minima) rotational bands.
 Interpretation was provided in terms of a (one dimensional) barrier
  with the third minimum and the third hump.
 Measurements of the angular distribution of the fission fragments
 at those excitation energies support this interpretation.
 Results for $^{232}$Th were consistent with a relatively shallow
 $~300$ keV minimum\cite{Blons1,Blons2,Blons3}. Later experiments claimed much deeper
 third minima in a number of nuclei, including $^{234,236}$U, see
 \cite{Krasznahorkay2011} and references therein.

\begin{figure}[h]
\vspace{-0mm}
\hspace{0mm}
\centerline{\includegraphics[scale=0.8]{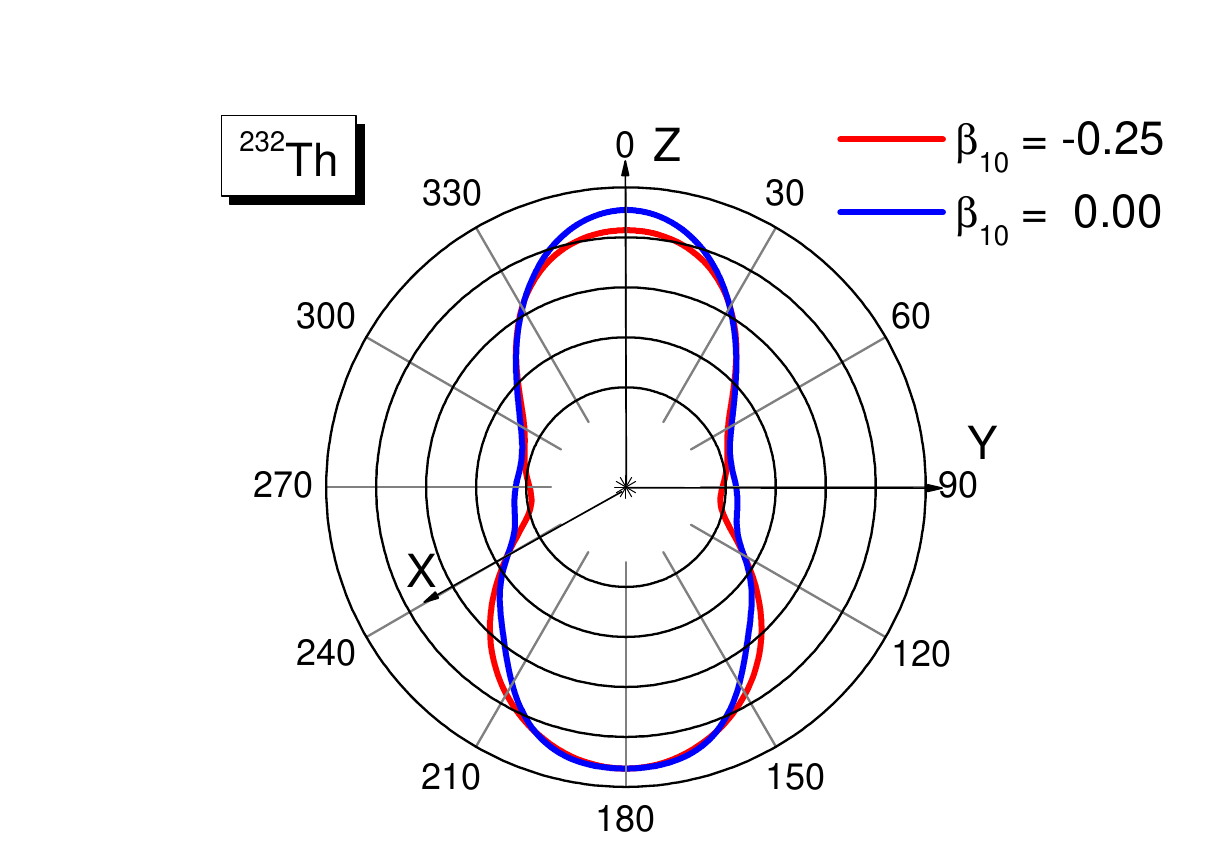}}
\vspace{-0mm}
\caption{{\protect Influence of the $\beta_1$ deformation on the nuclear
 shape. Taken from \cite{Jachimowicz2013}; copyright 2013 by APS.  }}
\label{shapes2}
\end{figure}

\begin{figure}[h]
\vspace{-0mm}
\hspace{4mm}
\centerline{\includegraphics[scale=0.70]{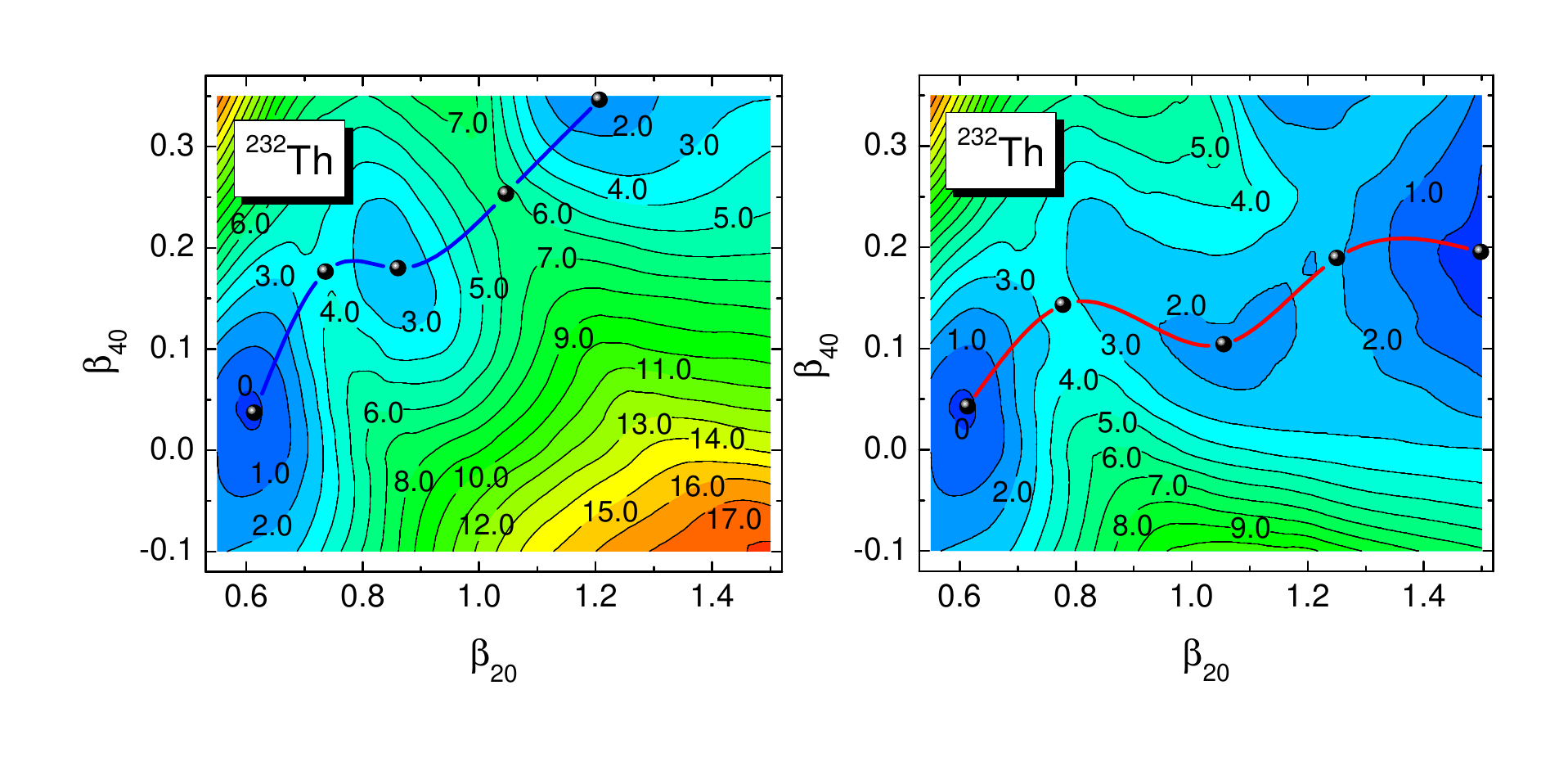}}
\vspace{-0mm}
\caption{{\protect Effect of $\beta_1$ on the energy surface of $^{232}$Th;
 two paths through the barrier correspond to shapes constructed by
 omitting (blue) and using (red) deformation $\beta_{10}$.
 Taken from \cite{Jachimowicz2013}; copyright 2013 by APS. }}
\label{PES2}
\end{figure}

\begin{figure}[h]
\vspace{-0mm}
\hspace{4mm}
\centerline{\includegraphics[scale=0.25]{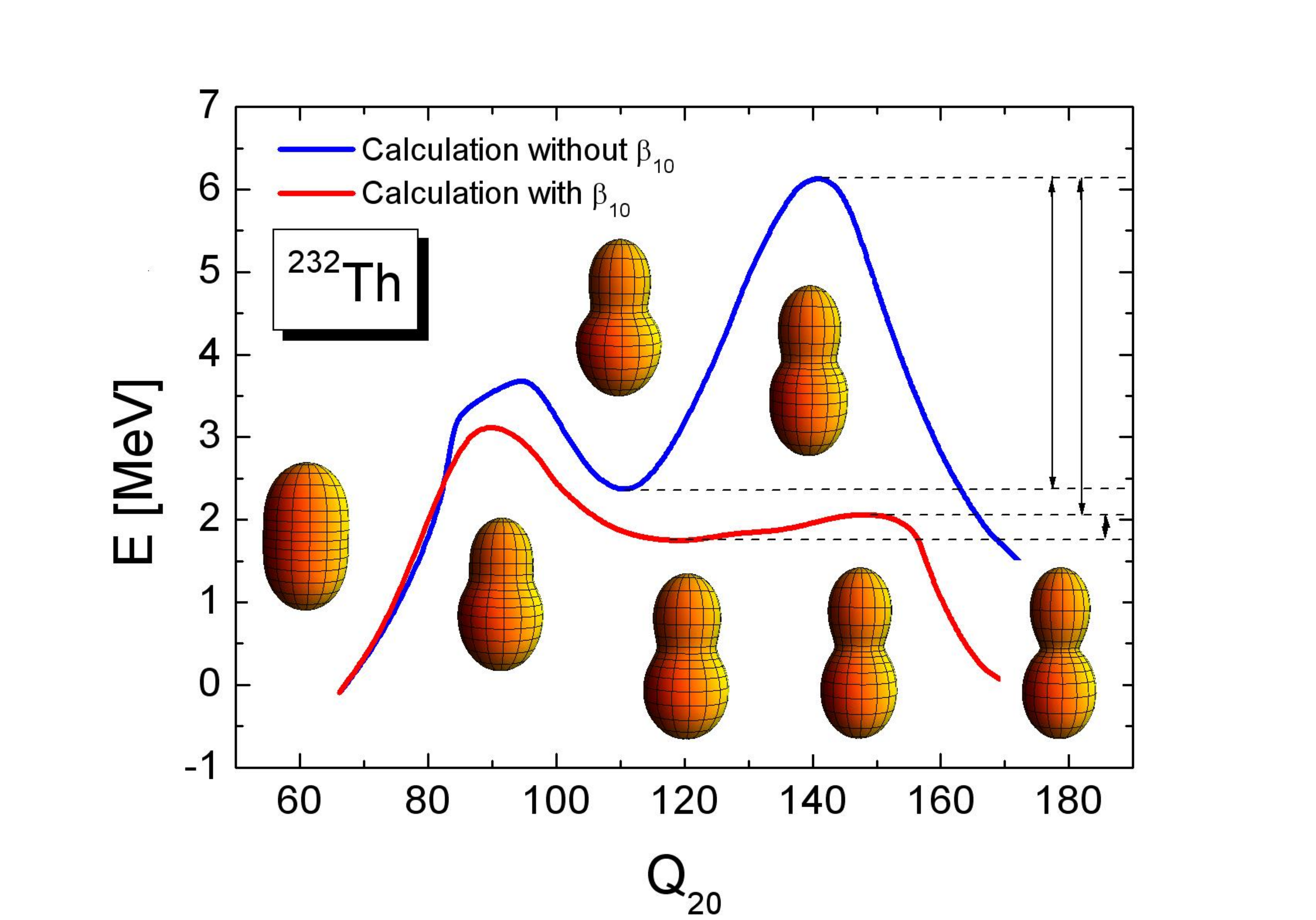}}
\vspace{-0mm}
\caption{{\protect Effect of $\beta_1$ on the effective one-dimensional barrier
 in $^{232}$Th shown as a function of quadrupole mass moment.
 Taken from \cite{Jachimowicz2013}; copyright 2013 by APS. }}
\label{Q2}
\end{figure}

  Theoretical selfconsistent calculations as well as earlier macro-micro
 results showed at most shallow third minima and, as shown in
 M. Kowal and J. Skalski \cite{KowSkal},
  and P.~Jachimowicz et al. \cite{Jachimowicz2013},
 this is also the result of the present model.
 These minima correspond to quite elongated, reflection - asymmetric shapes
 whose description in the parametrization (\ref{shape})
 already poses a problem.
 For their proper treatment it turned out necessary to explicitly include
 the dipole deformation $\beta_{10}$, which is spurious at small deformations
 for which it merely shifts the center of the shape.
 The effect of including $\beta_{10}$ is illustrated in Fig. \ref{shapes2},
 \ref{PES2}, \ref{Q2}. The two, similarly looking shapes differing only
 by $\beta_{10}$ shown in Fig. \ref{shapes2} correspond to quite different
 energies. The energy landscapes for $^{232}$Th in Fig. \ref{PES2}, obtained
 with and without $\beta_{10}$, show a drastic reduction of the
 third barrier by $\beta_{10}$. This is also seen in Fig. \ref{Q2}, in which
 the barriers along a possible fission paths are shown.

  It has to be emphasized that the effect of $\beta_{10}$ on the
 second barrier in actinides is weak, as documented in two
 calculations by P.~Jachimowicz et al.\cite{Jachimowicz2012,Jachimowicz2017_III}.



\subsection{Fission of $K$-isomers at the Second Minimum}

  One of the unsolved questions is how much of the symmetry of a given
 one-, two- or a few-quasiparticle configuration at the g.s. or secondary
 minimum is preserved along the barrier which controls its fission
 rate. This problem is relevant to fission rates of odd-$A$ and odd-odd nuclei
 as well as high-$K$ isomers.
 Nonaxial saddle deformations, like those at the first barriers
 in actinides, suggest that in their transmission the $K$ quantum number
 is probably not preserved.
 An interesting case are the possible high-$K$ isomers at the
 superdeformed second minima in actinides for which the axial symmetry of the
 second barrier would imply the conservation of the $K$ quantum number in
 fission. That could lead to a substantial increase in the barrier height
 for such isomers.
\begin{figure}[h]
\centerline{\includegraphics[scale=0.6]{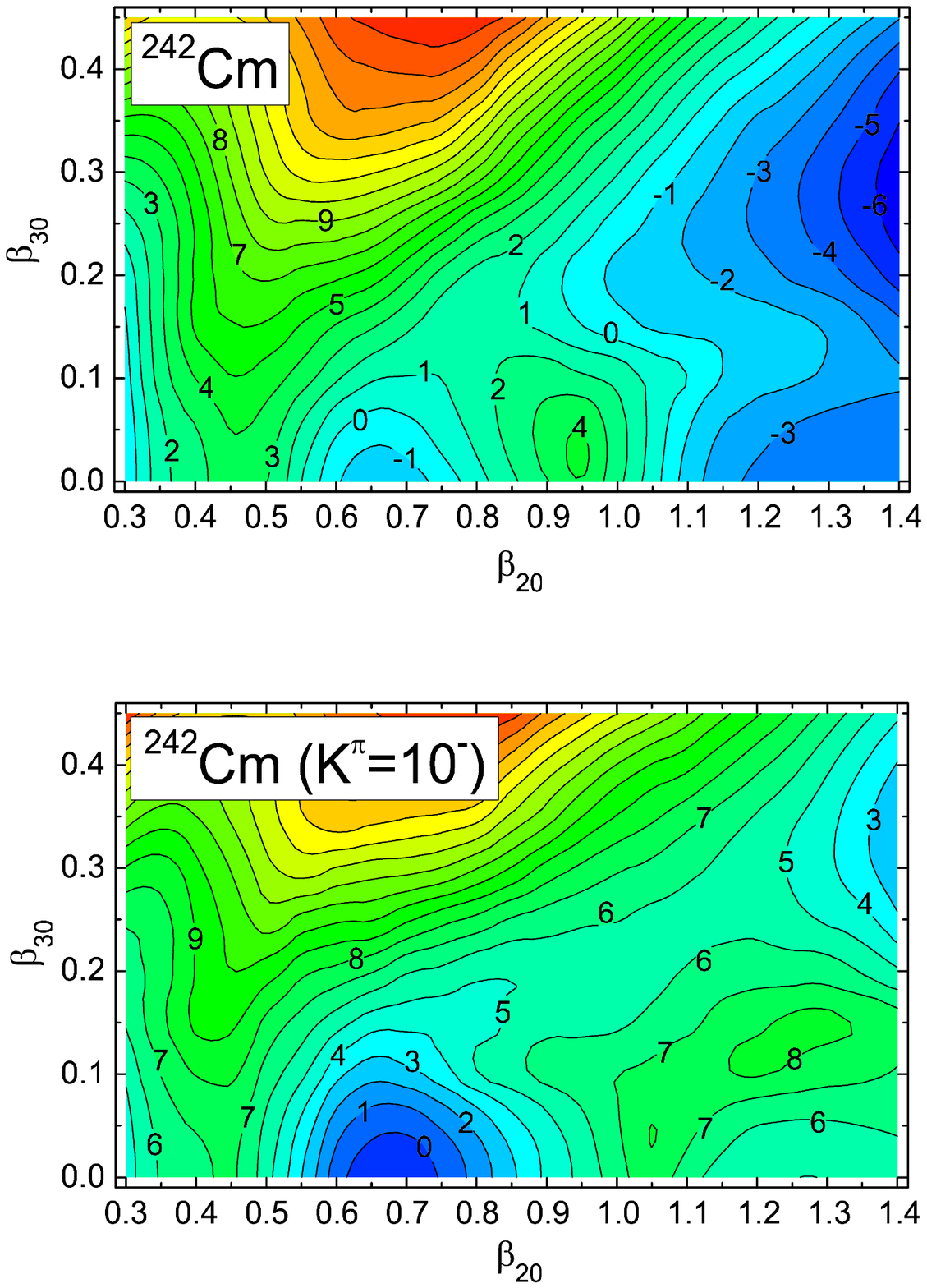}}
\caption{{\protect Energy relative to the macroscopic energy at the spherical
 shape in $(\beta_{20},\beta_{30})$ plane for the lowest and $K^{\pi}=10^-$
 configuration in $^{242}$Cm around and beyond the second minimum, obtained
 by minimization over remaining coordinates $\beta_{40}$-$\beta_{80}$. The
 effect of keeping configuration on the fission barrier may be clearly seen.
 Taken from \cite{Brodzinski2018}; copyright 2018 by APB. }}
\label{ISOMER242Cm}
\end{figure}

 In tables \cite{Singh2000} one finds that
 in each even-even $^{240-244}$Cm there were detected two
 isomeric fission half-lives differing by 3 - 4 orders of magnitude,
 with $T_1=40$ ps and $T_2=180$ ns in $^{242}$Cm. Energy landscapes calculated
 for this nucleus around and beyond the second minimum are shown in Fig.
 \ref{ISOMER242Cm} (taken from \cite{Brodzinski2018}) for the adiabatic (i.e. the lowest configuration
 at each deformation) and $K=10$ configuration, dominantly two-neutron
 $\nu 11/2^+,\nu 9/2^-$ in the second well, which, according to the level
 scheme of the W-S potential, is the main candidate for a $K$-isomer there.
 It may be seen that a large increase in the height ($\approx 4$ MeV) and
 width of the second fission barrier is predicted for the $K=10$ configuration.
 If the situation presented in Fig. \ref{ISOMER242Cm} is real, then the
 difference in fission half-lives $T_1$ and $T_2$ would rather come from a
 delay of the {\it electromagnetic} transition from the isomer to the g.s.
 in the second well (with the subsequent fission of the latter) than
 from the direct fission of the isomer, for which the increased fission
 barrier suggests a longer fission half-life.

  \section{Barriers in Superheavy Nuclei}

 While results in the actinide region can serve as a check the model's
 consistency with experimental data, calculations in the region of superheavy
 nuclei are predictions involving a serious extrapolation. One should bear
 in mind that this concerns all methods used, does not matter how 'realistic'
 or 'fundamental' they claim to be. The real issue at hand is the
 dependence of the mean-field or phenomenological potential on $Z$ and $N$ in
 the uncharted region of the table of isotopes. An uncertainty regarding this
 dependence is directly reflected in the uncertain next magic numbers beyond
 lead which come out differently in different models. Of course, issuing
 fission barriers are also strongly model - dependent, even if in
 the actinide region they were similar.

\begin{figure}[h!]
 \includegraphics[scale=0.35]{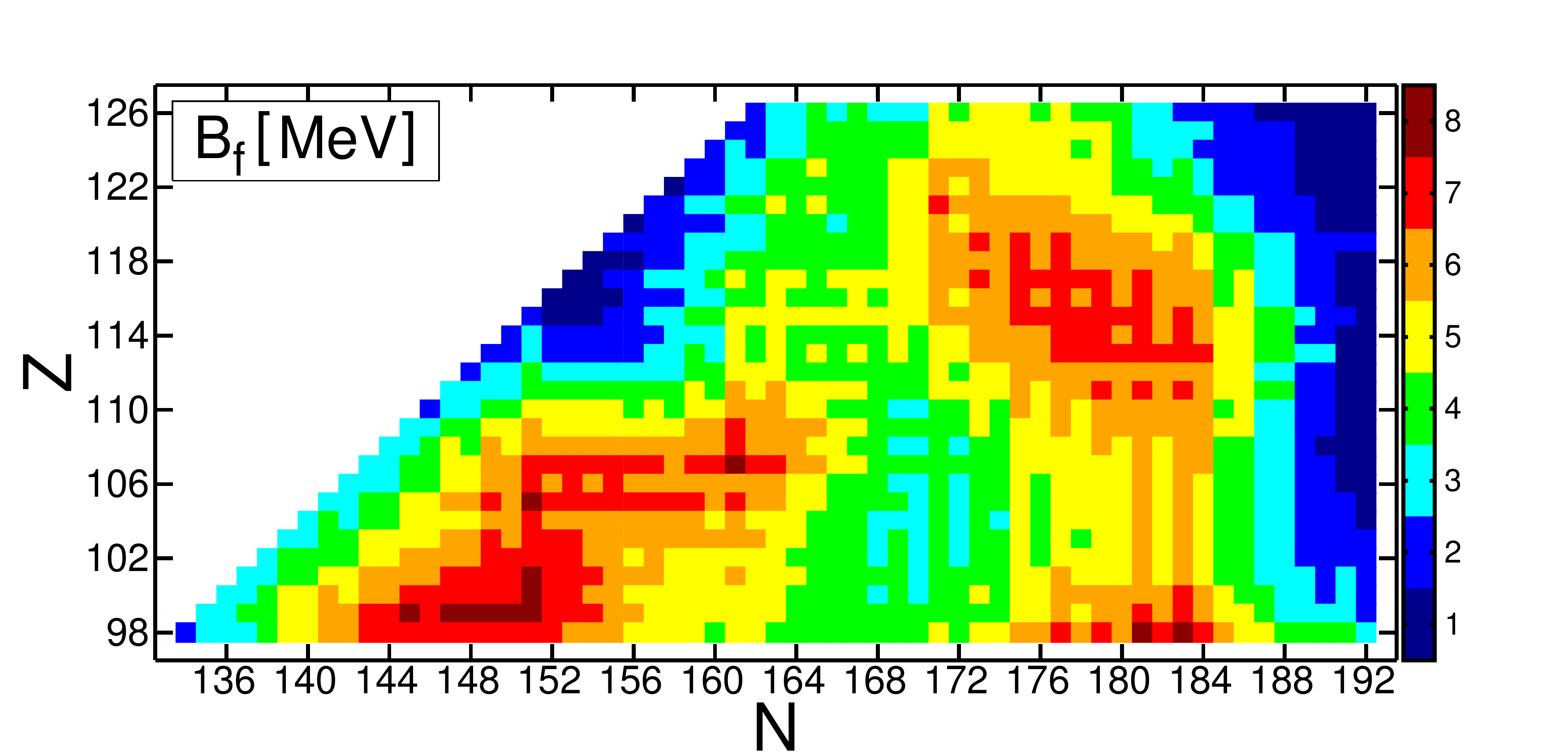}
\caption{Calculated fission barriers $B_{f}$ for superheavy nuclei according to MM model described in this Chapter.
Figure taken from \cite{Jachimowicz2017_I}; copyright 2017 by APS.}
\label{Bftot}
\end{figure}

All calculated fission barrier heights within the presented MM approach were
 collected and shown as a map $B_{f} (Z,N)$ in Fig. \ref{Bftot}.
One can see three areas of clearly raised fission barrier:
 ($Z\approx 102 , N\approx 152$),
($Z\approx 108,  N \approx 162$), ($Z\approx 114,  N\approx 180 $) and the
 region of low barriers around $N \approx 170$.
The effect of the odd particle, i.e., an often (but not always) higher
 barrier in the neighbouring odd-particle system, can be seen in
 Fig. \ref{Bftot}.
Looking at this global set of barriers in the superheavy nuclei, one can
 conclude that
in the whole region $Z=98-126$ the predicted fission barriers are limited,
 $B<8$MeV.
\begin{figure}[h!]
 \includegraphics[scale=0.35]{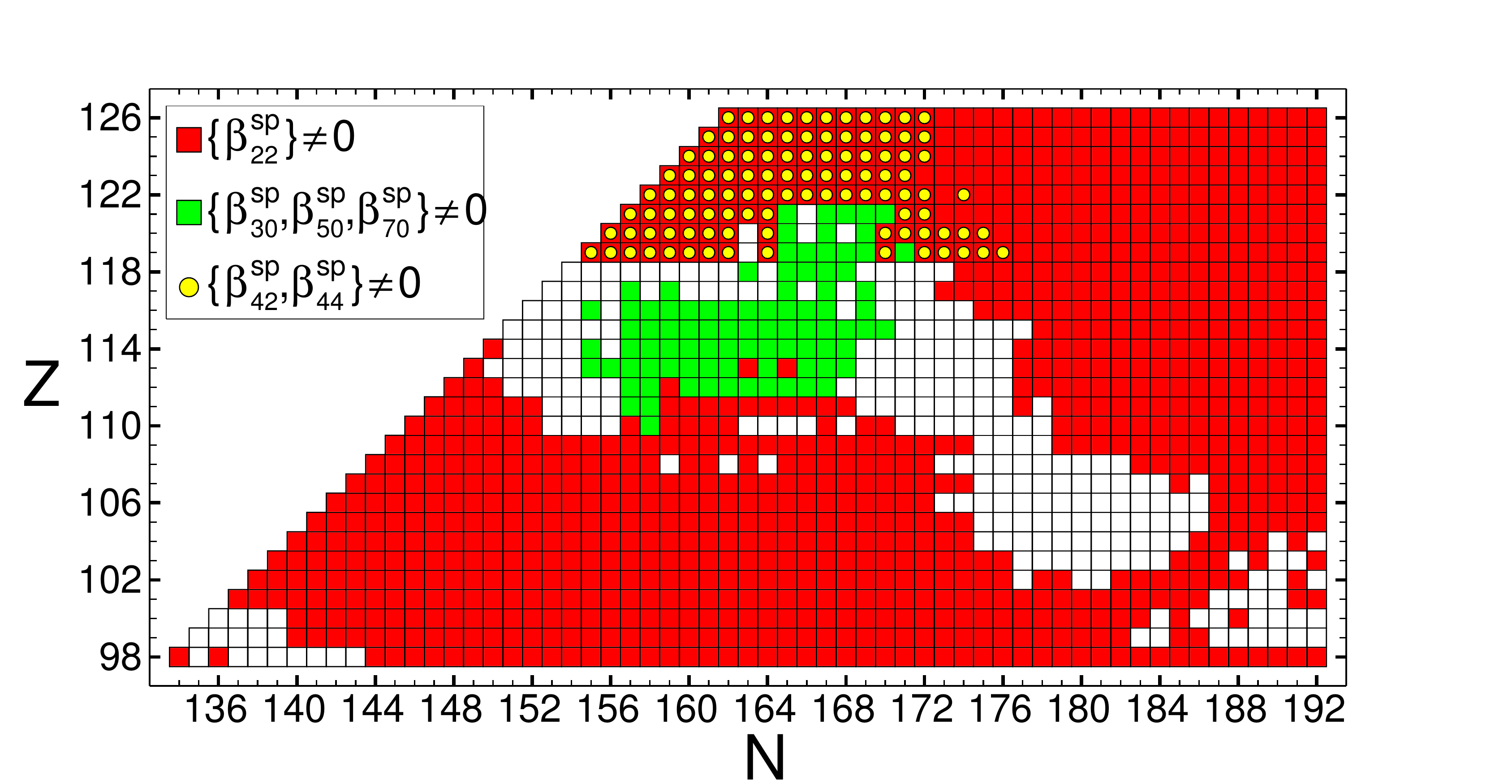}
 \caption{Nuclei with $\beta_{22}$ deformation at the saddle point (red squares),
mass-asymmetric fission saddle point (green squares) and saddles with
nonaxial hexadecapole deformation (yellow circles).
White regions in the considered area denotes nuclei with axially- and reflection- symmetric saddle point
\cite{Jach2021}; copyright 2021 by Elsevier.}
\label{Bfshapes}
\end{figure}

The categories of obtained saddle-point shapes are shown in Fig. \ref{Bfshapes}, taken from ref \cite {Jach2021}.
 As one can see, the saddles are mostly triaxial. The effect of the lowering
 of the axially-symmetric barrier by the quadrupole nonaxiality $\beta_{22}$,
 marked in red, is very systematic and affects almost 3/4 of all considered
 nuclei. The green color in Fig. \ref{Bfshapes} shows the mass-asymmetric
 saddle points. The nuclei having triaxial shapes at saddle points with
 additional hexadecapole nonaxiality $\beta_{42}$ and $\beta_{44}$ are marked
 by yellow dots.
Here belong many neutron deficient nuclei with $Z \geq 119$.
 \begin{figure}[h!]
 \includegraphics[scale=0.6]{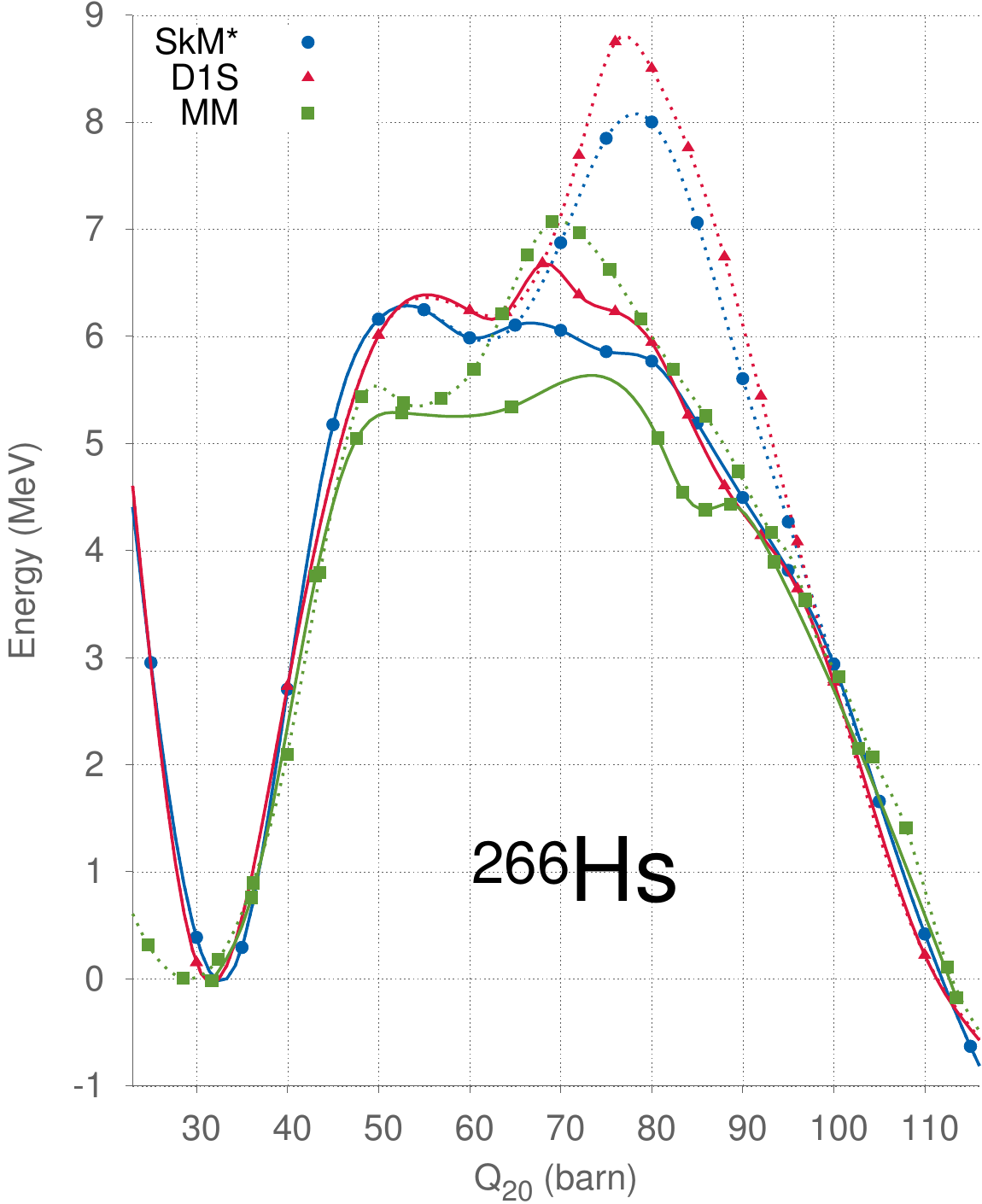}
 \caption{Fission barrier of $^{266}$Hs as function
  of the quadrupole moment $Q_{20}$ calculated in
  MM model (green, filled squares), Skyrme HFB
  approach (blue, filled circles) and Gogny HF model (red, filled
  triangles).  Dashed lines represent the raw axial  barrier.
   Solid lines correspond to the barriers with traxiality - $\gamma$.
   Taken from \cite{Baran}; copyright 2015 by Elsevier.}
 \label{BfHs}
\end{figure}

 For an additional insight, one can crosscheck barriers predicted
 within other models. We start with the comparison of the fission barriers for $^{266}$Hs
 calculated in the frame of various models based on different methodologies.
 The barriers derived in the Skyrme HFB \cite{SKM}, Gogny HF \cite{Gogny} and
 the present MM model
 are shown in Fig.~\ref{BfHs}, taken from \cite{Baran}, as a function of the
 quadrupole moment $Q_{20}$.
One can see that for this particular nuclide there is a relatively good
 agreement in barrier heights.
The differences in the peak height (the saddle point) reach 1 MeV.

\begin{figure}[h!]
 \includegraphics[scale=0.6]{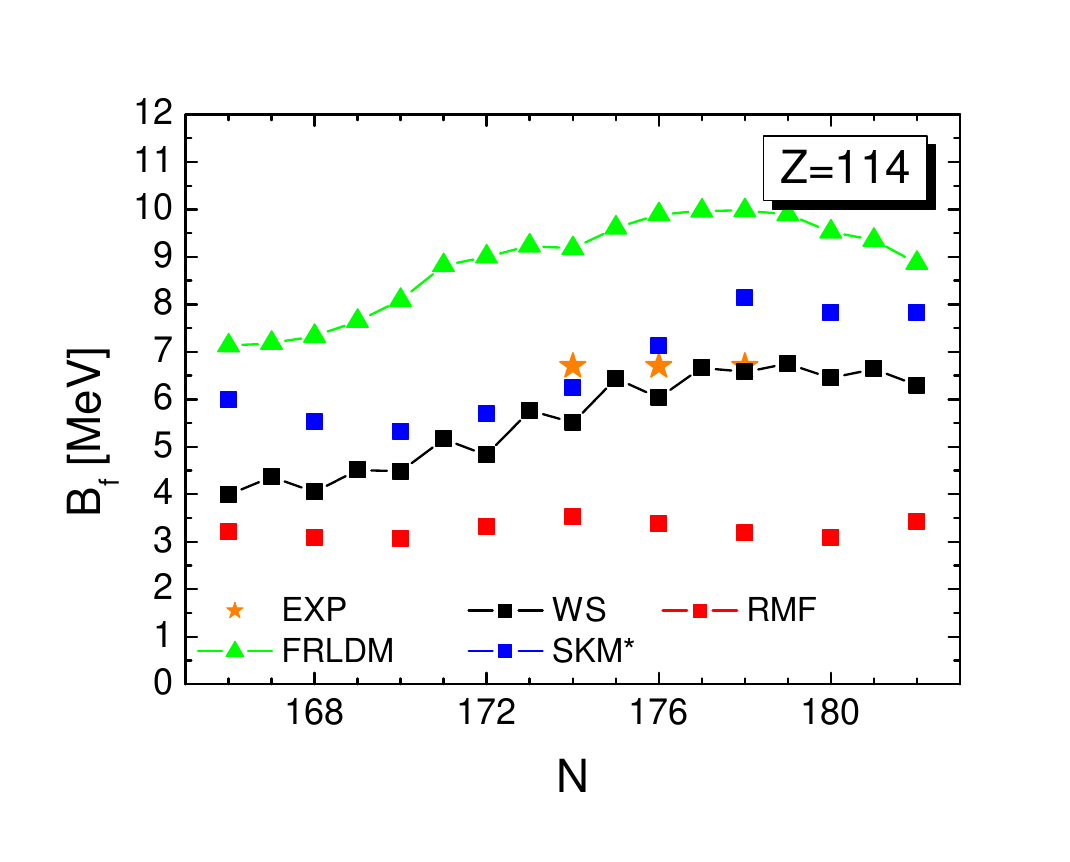}
  \includegraphics[scale=0.6]{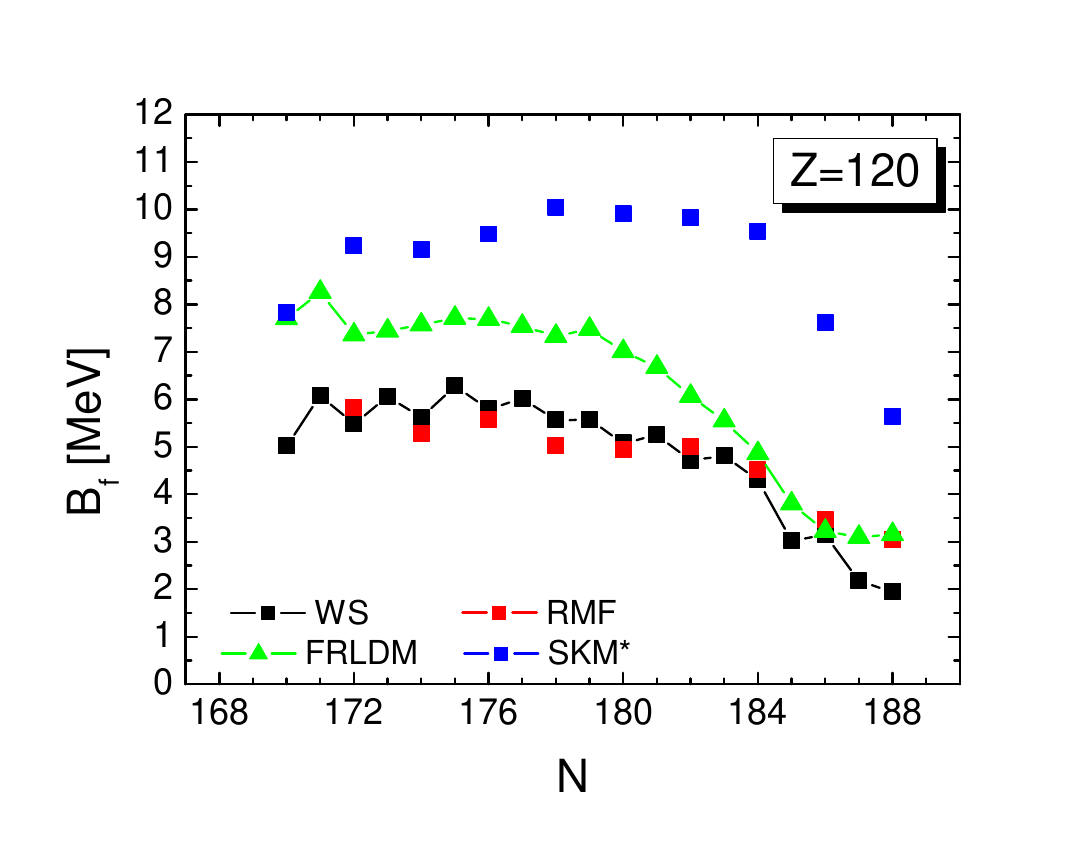}
 \caption{ Fission barriers predicted by various models: black - WS model, green – FRLDM \cite{FRLDM}, blue – SkM* \cite{SKM}, red – RMF with NL3 parametrization \cite{RMF} and experimental data taken from \cite{ITKIS} for Z=114 (left panel) and for Z=120 (right panel) isotopes. Taken from \cite{Jachimowicz2017_I}; copyright 2017 by APS.}
 \label{114th120th}
\end{figure}

The main conclusion from Fig. \ref{Bfshapes} and \ref{BfHs}  is that the
 effect of triaxiality is significant in all models and {\it should not}
 be ignored. Therefore, we have chosen for a more systematic
 comparison/discussion only the models
 that take this effect into account.
Furthermore, while all models predict similar barrier heights in the area of
 the element Hs, they differ considerably for much heavier nuclei.
It is clearly visible in Fig. \ref{114th120th} for nuclei with $Z = 114$ and
 $Z = 120$.

For Flerovium isotopes, the barriers calculated by the MM model and with the
 SKM* interaction \cite{SKM}
agree with the experimental (empirical) estimates \cite{ITKIS}.
 The FRLDM \cite{FRLDM} overestimates these
 quasi-empirical barriers \cite{ITKIS} significantly.
Although only the lower limit for the barrier height has been estimated in
 \cite{ITKIS}, which would reproduce the known fusion cross-sections at the
 picobarn level, such a high barrier seems problematic.
 On the other hand, with small barriers obtained within the RMF model,
 one cannot explain the experimentally known millisecond fission half-life
 in $^{284}$Fl. One should note, however, that a slight tuning of the RMF
 model \cite{RMF2015} gives higher barriers, closer to the MM results
 presented here.

 For $Z=$120 the MM results are very close to those
 obtained within the RMF model, see Fig. \ref{114th120th}.
 The results of \cite{FRLDM} are systematically higher by $\approx$ 1 MeV.
 This is in contrast with the Skyrme SkM* prediction \cite{SKM}
 of the high barriers for $Z=120$ \cite{SKM}, related to the proton magic gap.
 Three models: FRLDM, RMF and MM converge
 to $B_{f} \simeq 5$ MeV at N=182-184.
 The nucleus $^{302}$120 is particularly interesting, as two unsuccessful attempts to produce it have already taken place in GSI,
 providing a cross-section limit of 560 fb \cite{120GSI1} or 90 fb in
 \cite{120GSI2}, and in Dubna \cite{120DUBNA}, providing the limit of 400 fb.
 The cross-section estimates \cite{Wilczynska3} do not support a possibility
 of easy production of this SH isotope in the laboratory.
 However, with the barrier of the order of 10 MeV, as obtained in the
 frame of the self-consistent Skyrme SkM* theory,
 producing superheavy Z=120 nuclei, it seems,
 should not pose any difficulties.

\section{Conclusions}

 Nuclear fission is an extremely complex process, difficult
 to analyse within a genuinely microscopic quantum theory.
 In order to gain a qualitative understanding of it and obtain
  quantitative estimates of experimentally measured observables
 we use a number of concepts, like the nuclear deformation and
  the fission barrier or, more generally, energy landscape.
  Although not directly measurable, they provide clues about physics
 of fission, are ingredients in formulas for probabilities of various
 reactions and decays and give hints concerning the synthesis and
 detection of the heaviest nuclei.

  The ingenious method of shell correction by Strutinsky allowed
 to bypass severe limitations of microscopic theories and calculate
   in a relatively simple manner energy landscapes of nuclei with an
  acceptable error.
  Although large selfconsistent mean-field calculations are
  possible today, the macro-micro method is still competitive in areas
 like nuclear fission, in which it is necessary to account for
  all important deformations in hundreds or more nuclei, including odd and
  odd-odd, which multiplies the numerical effort.

  The presented results of the macro-micro model illustrate the status of
   agreement between the calculated and evaluated from experiment
  fission barriers in 72 actinides, in which they are known.
  In view of the uncertainties, both in experimental evaluations and in
  the macro-micro method, the obtained agreement seems reasonable.
  A slightly worse agreement was obtained in the FRLDM model, which
  encompasses much larger number of nuclei, while the selfconsistent
  results in actinides are of very mixed quality, and nearly always missing
  the odd and odd-odd nuclei.

  The calculated fission barriers in superheavy nuclei constitute a
 predictive part of the present chapter. They are compared to
  the results of other calculations, including the selfconsistent ones,
 and the differences between models are shortly discussed.
   These differences are strongly related to our ignorance as to the
   next magic numbers beyond lead, especially the one for protons.
   The large differences in predictions mean that experiment will
   eliminate some of the models someday.

%
%
%
%

\end{document}